%% file: sn-article.tex
\documentclass[pdflatex,sn-mathphys-num]{sn-jnl}

\usepackage{graphicx}%
\usepackage{multirow}%
\usepackage{amsmath,amssymb,amsfonts}%
\usepackage{amsthm}%
\usepackage{mathrsfs}%
\usepackage[title]{appendix}%
\usepackage{xcolor}%
\usepackage{textcomp}%
\usepackage{manyfoot}%
\usepackage{booktabs}%
\usepackage{algorithm}%
\usepackage{algorithmicx}%
\usepackage{algpseudocode}%
\usepackage{listings}%
\usepackage{comment}

\theoremstyle{thmstyleone}%
\theoremstyle{thmstyletwo}%

\theoremstyle{thmstylethree}%

\usepackage{rh_defs_21}

\newcommand{\kspace}[1]{\textit{k}-space {#1}}

\begin{document}


\title{Deep learning of personalized priors from past MRI scans enables fast, quality-enhanced point-of-care MRI with low-cost systems}




\author[1,2]{\fnm{Tal} \sur{Oved}}\email{tal.oved@campus.technion.ac.il}

\author[3]{\fnm{Beatrice} \sur{Lena}}\email{b.lena@lumc.nl}
\equalcont{These authors contributed equally to this work.}

\author[3]{\fnm{Chloé} \sur{F. Najac}}\email{c.f.najac@lumc.nl}
\equalcont{These authors contributed equally to this work.}

\author[4]{\fnm{Sheng} \sur{Shen}}\email{sshen6@mgh.harvard.edu}

\author[4]{\fnm{Matthew} \sur{S. Rosen}}\email{msrosen@mgh.harvard.edu}

\author[3]{\fnm{Andrew} \sur{Webb}}\email{a.webb@lumc.nl}

\author*[1,2,5]{\fnm{Efrat} \sur{Shimron}}\email{efrat.s@technion.ac.il}

\affil[1]{%
  \orgdiv{Department of Electrical and Computer Engineering}, 
  \orgname{Technion – Israel Institute of Technology}, 
  \orgaddress{%
    \city{Haifa}, 
    \postcode{3200004}, 
    \country{Israel}}}

 \affil[2]{\orgdiv{May-Blum-Dahl Technion Human MRI Research Center}, \orgname{Technion - Israel Institute of Technology}, \orgaddress{ 
\city{Haifa}, \postcode{3200004}, 
 \country{Israel}}}

\affil[3]{%
  \orgdiv{Department of Radiology}, 
  \orgname{C.J. Gorter MRI Center, Leiden University Medical Center}, 
  \orgaddress{%
    \street{Albinusdreef 2}, 
    \city{Leiden}, 
    \postcode{2333 ZA}, 
    \country{The Netherlands}}}

\affil[4]{%
  \orgdiv{Department of Radiology}, 
  \orgname{Athinoula A. Martinos Center for Biomedical Imaging, Massachusetts General Hospital and Harvard Medical School}, 
  \orgaddress{%
    \street{149 13th Street}, 
    \city{Charlestown}, 
    \state{MA}, 
    \postcode{02129}, 
    \country{USA}}}

\affil[5]{%
  \orgdiv{Department of Biomedical Engineering}, 
  \orgname{Technion – Israel Institute of Technology}, 
  \orgaddress{%
    \city{Haifa}, 
    \postcode{3200004}, 
    \country{Israel}}}


\abstract{Magnetic resonance imaging (MRI) offers superb-quality images, but its accessibility is limited by high costs, posing challenges for patients requiring longitudinal care. Low-field MRI provides affordable imaging with low-cost devices but is hindered by long scans and degraded image quality, including low signal-to-noise ratio (SNR) and tissue contrast. We propose a novel healthcare paradigm: using deep learning to extract personalized features from past standard high-field MRI scans and harnessing them to enable accelerated, enhanced-quality follow-up scans with low-cost systems. To overcome the SNR and contrast differences, we introduce ViT-Fuser, a feature-fusion vision transformer that learns features from past scans, e.g. those stored in standard DICOM CDs. We show that \textit{a single prior scan is sufficient}, and this scan can come from various MRI vendors, field strengths, and pulse sequences. Experiments with four datasets, including glioblastoma data, low-field ($50mT$), and ultra-low-field ($6.5mT$) data, demonstrate that ViT-Fuser outperforms state-of-the-art methods, providing enhanced-quality images from accelerated low-field scans, with robustness to out-of-distribution data. Our freely available framework thus enables rapid, diagnostic-quality, low-cost imaging for wide healthcare applications.}



\maketitle


\section{Introduction}\label{sec1}
Magnetic resonance imaging (MRI) is a cornerstone of modern diagnostics, offering safe, radiation-free scans with exceptional image quality. However, conventional high-field MRI systems - typically operating at $1.5T$ or $3T$ - are extremely costly, with prices estimated at approximately $\$1$ million per Tesla. These systems also require high-end, costly infrastructure, hence, they are commonly located in hospitals or other highly equipped facilities. Therefore, MRI scans commonly involve high costs and long wait lines, and their availability is quite limited \cite{arnold2023low,geethanath2019accessible}. This poses a significant challenge for vulnerable populations that need ongoing monitoring, such as cancer patients receiving treatment, for whom repeated visits to clinical facilities can be physically challenging and increase the risk of complications.


Low-field MRI is an emerging technology that provides compact, portable systems, suitable for point-of-care imaging \cite{sarracanie2015low, marques2019low, cooley2021portable, liu2021low, campbell2023low, Kimberly2023, tian2024new}, hence it can substantially increase the affordability and accessibility of MRI scans. Low-field systems operate at magnetic field strengths typically in the range of $0.01 T$ to $\sim 0.5T$, i.e., much lower than standard clinical MRI systems. Moreover, modern low-field systems are built from low-cost hardware, hence they are highly cost-effective - some models are estimated to cost only about  $\$20,000$ to  $\$200,000$ \cite{liu2021low,obungoloch2023site}. They can also run on standard power outlets without requiring dedicated infrastructure. This technology hence enables doing MRI scans in previously inaccessible settings, including at the patient's bedside \cite{Kimberly2023}, in mobile units \cite{deoni2022development},  outdoors, and at home \cite{guallart2022portable}. Therefore, it offers substantial potential to transform healthcare services by reducing the cost of MRI, increasing availability, and enabling more flexible patient-centered workflows.

However, the clinical adoption of low-field MRI is currently limited by critical barriers: low image quality - due to low signal-to-noise ratio (SNR), low resolution, reduced tissue contrast, and prohibitively long scan durations. In particular, the reduced SNR poses substantial challenges. In MRI, the SNR depends on the magnetic field strength, and shifting from the standard $1.5–3T$ regime to the sub-tesla regime reduces the SNR by several \textit{orders of magnitude} \cite{hoult1976signal,marques2019low}. Moreover, low-field systems commonly operate without infrastructure, i.e. without standard shielding from electromagnetic noise, hence the SNR is further reduced. To mitigate this, low-field data are commonly acquired in 3D, as volumetric imaging offers better SNR than 2D imaging, and scans are often repeated to reduce noise through data averaging. However, these common practices come at the cost of very long scan durations \cite{marques2019low,geethanath2019accessible,shimron2024accelerating,man2023deep,tian2024new}. Another major limitation is the low tissue contrast. Because tissue-relaxation parameters (e.g. T1, T2) vary with the the magnetic field strengh, some pulse sequences that are standard in high-field MRI produce images with reduced contrast at low fields \cite{campbell2023low}. Therefore, there are unmet needs for new techniques to improve scan speed and image quality in low-field MRI. 

Compressed sensing and deep learning (DL) have been extensively studied for 
acceleration and image reconstruction in high-field MRI \cite{Lustig2007,geethanath2013compressed,shimron2020temporal,zhu2018image,Hammernik2018,Aggarwal2019,Heckel2024,hammernik2023physics,Sharma2022,shimron2023ai}, and recent studies have begun to explore their use for low-field MRI 
\cite{man2023deep, shimron2024, Lau2023, koonjoo2021boosting, Zhao2024, Lin2023, Arefeen2024, deLeeuwdenBouter2022, Ayde2024, campbell2023low, tian2024new, kofler2025mr}. The clinical utility of low-field MRI has also been explored, often by comparing images acquired with low- and high-field systems \cite{Liu2021, campbell2023low, Mazurek2021, Kimberly2023, Sorby-Adams2024, tian2024new, Shen2024}. However, a common limitation of this body of work is that studies traditionally consider low-field and high-field MRI as an \textit{alternative} clinical options. To our knowledge, the integration of these modalities into more comprehensive workflows has not yet been considered.

Here, we propose a novel healthcare paradigm: treating low-field and high-field MRI as \textit{complementary} clinical modalities and developing new workflows that leverage their \textit{synergistic strengths}. Notably, these modalities offer complementary attributes: high-field MRI enables superior image quality but is associated with high costs and limited accessibility, whereas low-field MRI offers greater affordability and portability, albeit with reduced image quality and longer scan durations. By combining these technologies, we aim to create integrated clinical workflows that better serve patients who need frequent monitoring. Moreover, we aim to utilize the high-quality data offered by high-field systems to accelerate and improve the image quality in low-field scans.

\begin{figure}[htbp]
\centering
\includegraphics[width=0.99\textwidth]{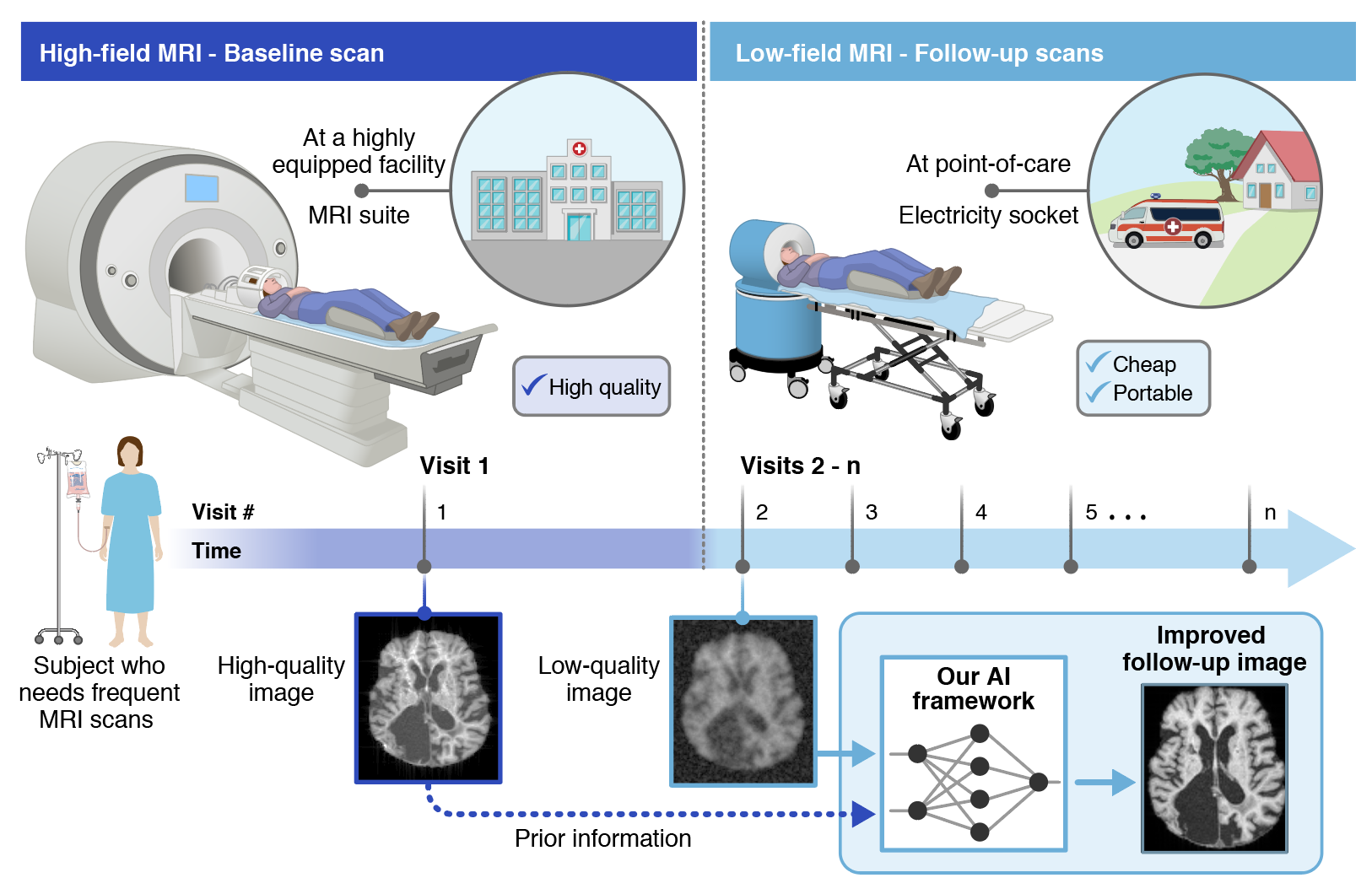}
\caption{\textbf{Overview of our proposed personalized imaging paradigm combining high-field and low-field MRI.} We introduce a novel clinical workflow where a high-quality baseline scan is acquired at a highly equipped facility using a standard high-field MRI system (e.g., $1.5T$ or $3T$), and is followed by repeated, accessible low-field scans (e.g., $47mT$) at the point of care. While low-field MRI offers low-cost, portable imaging, it suffers from long scan durations and degraded image quality due to low SNR and reduced tissue contrast. Our approach leverages prior information from the baseline high-field scan to enhance follow-up low-field scans using a deep learning model, ViT-Fuser. This AI framework extracts personalized features from the prior scan and integrates them during image reconstruction, enabling accelerated, diagnostic-quality follow-up scans. The paradigm enables frequent, high-quality monitoring for patients requiring longitudinal care, without repeated hospital visits.}\label{intro}
\end{figure}

Specifically, we propose a new clinical workflow that begins with a high-field MRI scan, which is standard for diagnostic evaluations or surveillance imaging, and - unlike current practices - continues with follow-up low-field scans at the point of care, using compact portable systems (Fig. \ref{intro}). This synergistic approach combines the diagnostic power of a high-quality baseline scan with the convenience and accessibility of portable follow-up scans, hence it can make healthcare services more available and reduce the need for patients to make repeated visits to major clinical centers.


Furthermore, we propose a new \textit{computational strategy}: using DL to learn personalized information from the initial high-field scan and apply it to accelerate and enhance image quality of the follow-up low-field scans. This approach can efficiently harness the complementary field strengths of these two modalities and thus enable more efficient and accessible imaging pipelines. To the best of our knowledge, this approach has not been explored,  perhaps because it involves challenges; it requires addressing the inherent differences between low- and high-field data, particularly differences in SNR and tissue contrast, and any potential longitudinal anatomical changes.


\begin{figure}[htbp]
\centering
\includegraphics[width=0.99\textwidth]{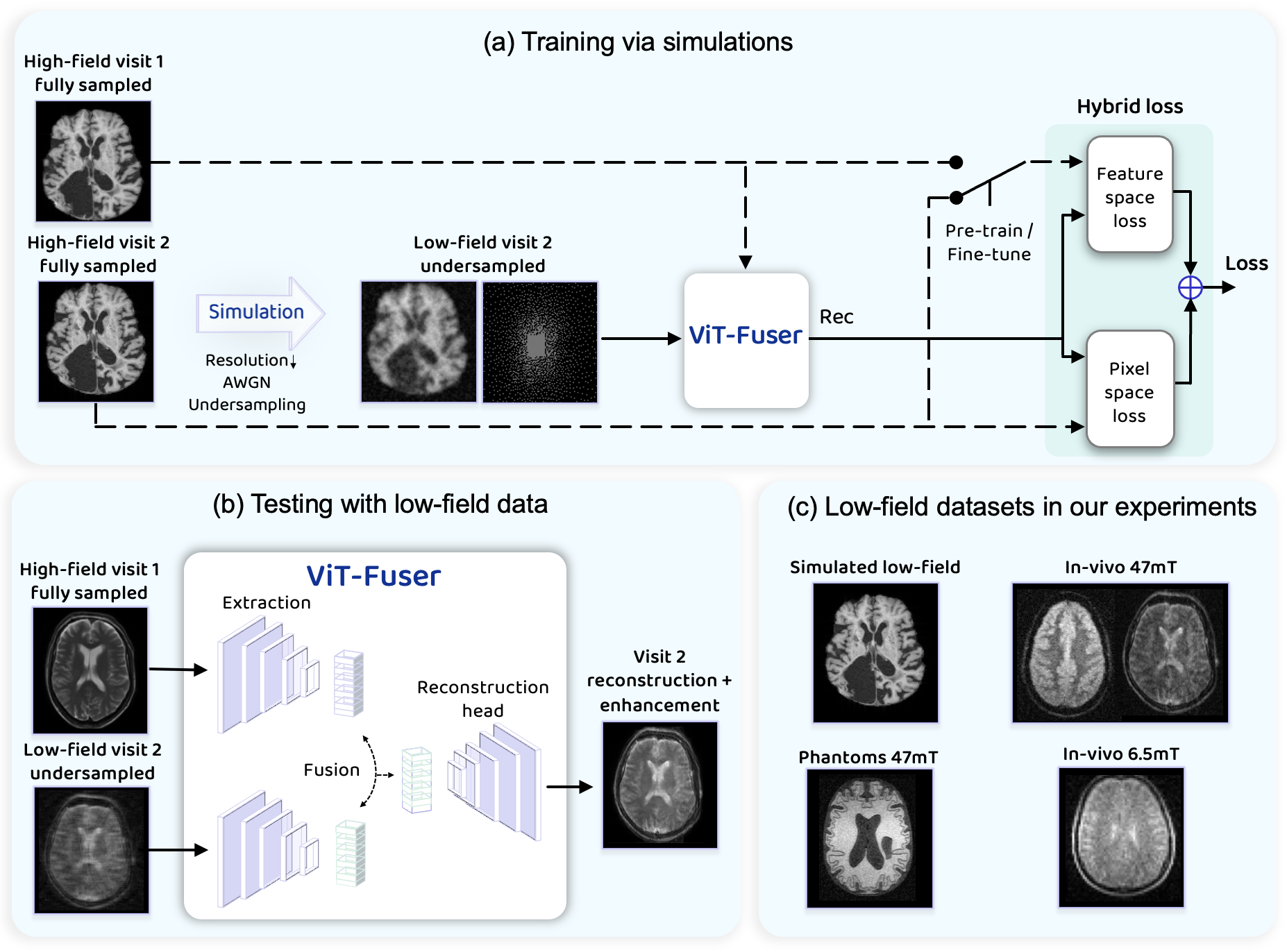}
\caption{\textbf{Our framework and training strategy for the proposed multi-field-strengh clinical workflow.} Our training scheme (a) involves a pre-training stage via a simulation environment, utilizing longitudinal high-field data, overcoming the lack of low-field data. Additionally, we illustrate the proposed hybrid loss that combines both pixel space loss and feature space loss, which is calculated differently per training stage. We also demonstrate the inference process (b), where features extracted from a prior high-field MRI scan are leveraged to enhance the reconstruction of a current low-field scan. The diagram depicts the feature extraction, fusion, and reconstruction stages. Finally, we provide an overview of the different datasets of our experiments (c).}\label{setup}
\end{figure}

To enable materializing our multi-field-strength strategy, we introduce \textit{ViT-Fuser}, a DL framework that extracts personalized features from past standard high-field scans and incorporates them during image reconstruction in accelerated follow-up low-field scans. This framework consists of a vision transformer (ViT), feature fusion and reconstruction blocks (Fig. \ref{setup}). Notably, our framework can efficiently process and utilize data acquired from various MRI systems of different field strengths and vendors, and learn personalized features flexibly, without requiring perfect matching of the pulse sequences (i.e. tissue contrast) of the low- and high-field scans. This is demonstrated in an extensive set of experiments with several datasets.

Importantly, this work demonstrates that features extracted from \textit{a single prior high-field scan} are sufficient to both accelerate and enhance image quality in follow-up scans. To the best of our knowledge, this capability has not been previously shown. Moreover, we demonstrate that ViT-Fuser can efficiently learn personalized features from data stored in standard DICOM files, such as those commonly provided to patients on CDs by clinical vendors, hence it is highly practical and compatible with existing clinical workflows. Furthermore, ViT-Fuser is also highly robust to variations in SNR, acceleration factors, pulse sequences, image contrasts, and field strengths. 

This work hence introduces a novel healthcare paradigm and a highly general computational framework, widely suitable to a broad spectrum of systems and applications. To enable its exploration by the research community, upon publication, we will publicly release our codes and best-performing models.



\section{Methods}\label{sec2}


We begin with a brief description of our experimental setup, for clarity. More details are provided in section \ref{sec3}.

\textbf{\textit{Background}. } A challenge in developing the proposed ViT-Fuser model was the lack of public datasets that contain paired, longitudinal high- and low-field data, as this healthcare paradigm is proposed here for the first time. To address this, we trained the model through simulations with a longitudinal high-field database \cite{Suter2022} (Fig. \ref{setup} a). Later, we investigated the model's performance using four datasets (Fig. \ref{setup}b-c). In all cases, the prior data were from high field ($1.5T$ or $3T$) scans, while the follow-up data were from (i) simulations; (ii) low-field ($47mT$) phantom scans; (iii) in-vivo low-field ($47mT$) brain scans; (iv) and an ultra-low-field ($6.5mT$) brain data.  We note that this work substantially extends our published preliminary results, which included only simulations \cite{Oved2025}.

The ViT-Fuser model that was trained on the simulated data was used in all our experiments, and we refer to it as the pre-trained model. In the experiments with the two $47mT$ datasets, we fine-tuned this model using small amounts (a few slices) of such low-field data, to bridge the gap between the simulated and raw, low-field data, mainly changes in SNR and tissue contrast. In the experiment with the 6.5mT data, however, we did not  fine-tune the model, to examine its robustness to out-of-distribution data. 


\textbf{\textit{Training.}} We trained our model using the LUMIERE high-field ($1.5T$) longitudinal MRI database \cite{Suter2022} that contains scans of Glioblastoma (brain cancer) patients. From this dataset, we extracted 300 image pairs from 12 subjects, each consisting of an initial scan and a follow-up scan. For each pair, data from the first scan were generally used without changes, to represent initial high-field baseline scans. The only change made was a reduction of the resolution to $1.5$×$1.5 mm$ in-plane, to match that of follow-up low-field scans, which usually has reduced resolution. The second visit's data were artificially degraded to simulate an accelerated low-field acquisition. This included adding white Gaussian noise, reducing the resolution similarly to the first scan, and applying k-space (Fourier domain) undersampling. Next, every pair of images was registered using a standard registration algorithm \cite{Leutenegger2011, Fischler1987, MathWorksCVT2024}. 

During training, we used the original high-field data of the first visit as prior, and the degraded data of the second visit as the "acquired" low-field data. Both datasets were fed to the network as inputs (Fig. \ref{intro}). In addition, to enable our network to learn both reconstruction and image enhancement, we leveraged the availability of the high-field data of the second visit; we used these data as the training targets. This approach enabled our model to learn reconstruction, denoising, and image enhancement simultaneously. We emphasize that this is not a "data crime" \cite{Shimron2022} because this process was applied only during training, not during test time; it is more similar to a data augmentation technique. The training was done with our proposed \textit{hybrid loss}, which contains both pixel-space and feature-space computations (Fig. \ref{setup}a). Specifically, the feature-space loss computes the $l_1$ norm between the GRAM matrices of the feature maps of the two input images, allowing the feature distribution to be approximated through a simple matrix multiplication. 

\textbf{\textit{ViT-Fuser architecture}. }
Our transformer-based architecture consists of two heads that extract features from the high- and low-field data, a feature fusion block, and a reconstruction head (Fig. \ref{setup}b). It is designed to recover high-quality images from the undersampled low-field data
while integrating relevant features from a prior high-field scan and overcoming the inherent differences between high- and low-field data. This general approach can be easily adapted to various scenarios, including MRI systems by different vendors or various field strengths. 

\textbf{\textit{Evaluation.}}
We compared the performance of our ViT-Fuser to two well-established DL-based MRI reconstruction methods: MoDL \cite{Aggarwal2019}, which is an unrolled physics-guided method, and a state-of-the-art ViT \cite{Lin2022}. The three methods (MoDL, ViT, and ViT-Fuser) were trained and tested on the same datasets, and the hyperparameters of each method were calibrated for each database separately, to ensure optimal performance. We evaluated the performance using several standard image quality metrics, including SSIM \cite{Wang2004} and LPIPS \cite{Zhang2018}, and the state-of-the-art CMMD metric, which is highly correlated with human visual perception \cite{Jayasumana2023}.

\section{Results}\label{sec3}
\subsection{Longitudinal Glioblastoma data}

In the first set of experiments, we investigated the performance of our model using the LUMIERE database \cite{Suter2022}, where the follow-up images were degraded and then undersampled in k-space to mimic accelerated low-field MRI acquisitions. 

\begin{figure}[htbp]
\centering
\includegraphics[width=0.99\textwidth]{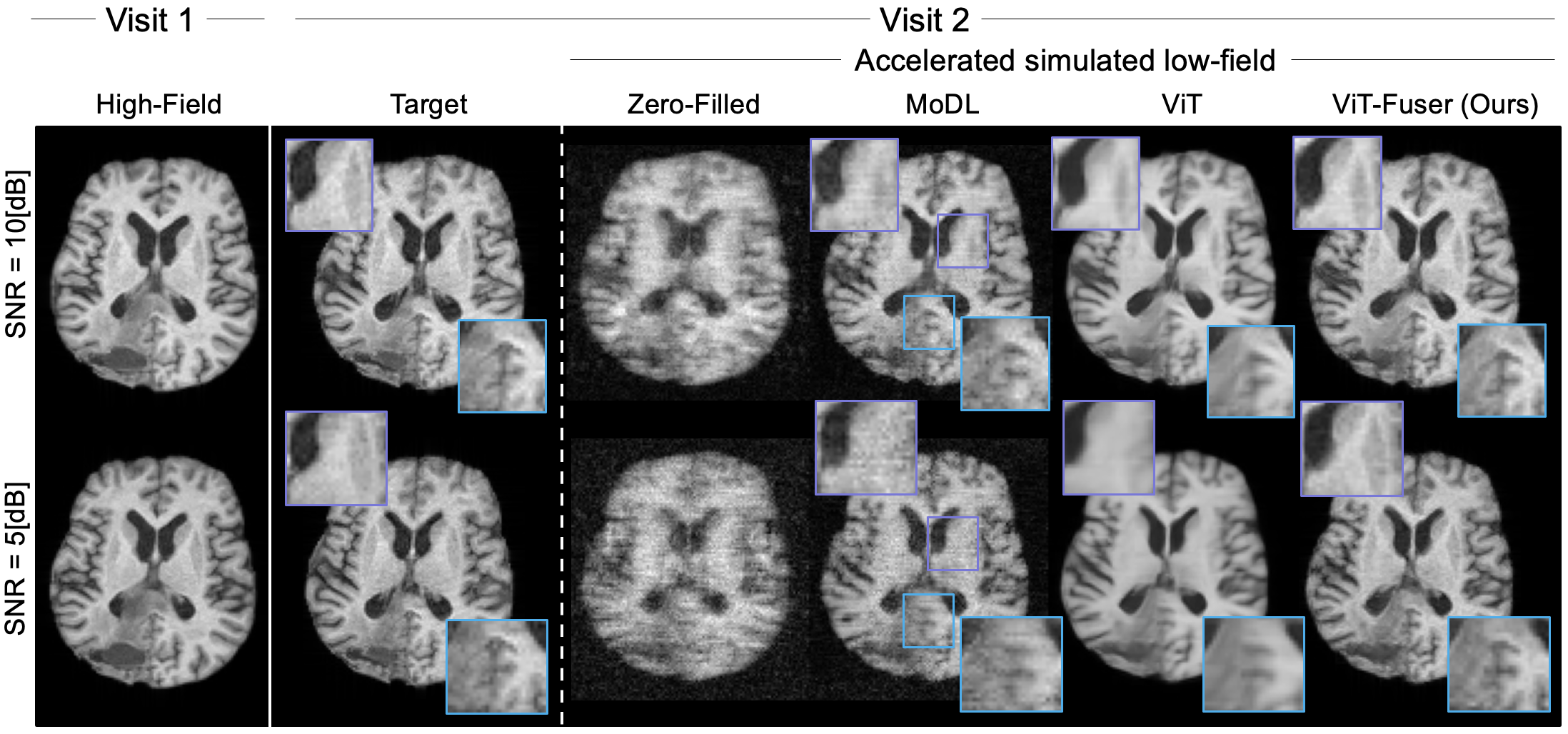}
\caption{\textbf{ViT-Fuser reconstruction compared to other methods: Accuracy and sharpness preservation for different SNR settings.} Comparison of MoDL \cite{Aggarwal2019}, ViT \cite{Lin2022}, and ViT-Fuser (ours) for accelerated imaging for in-vivo Glioblastoma data \cite{Suter2022} with 3-fold equispaced undersampling at SNR levels of $5dB$ (bottom row) and $10dB$ (top row). ViT-Fuser achieves improved fine-detail reconstruction and better texture preservation as SNR decreases from $10dB$ to $5dB$. These results indicate that using priors could be beneficial for improving image quality in low SNR regimes.
}\label{figequsipaced}
\end{figure}

\textbf{\textit{Robustness to various undersampling schemes and anatomical changes}}. Here we examined the model's performance for two undersampling schemes - equispaced (periodic) and variable-density undersampling. First, we performed experiments with equispaced sampling using an acceleration factor of $R=3$. In these experiments, the low-field MRI datasets were simulated with SNR levels of  $5dB$ or $10dB$. The results (Fig. \ref{figequsipaced}) indicate that ViT-Fuser provides reconstructions with higher quality than the other two methods. Specifically, it can be seen that MoDL provides a noisy image, while the ViT method removes the noise, but at the cost of over-smoothing and loss of small details. In contrast, our ViT-Fuser removes most of the noise while also providing sharp, anatomically accurate images.

\begin{figure}[htbp]
\centering
\includegraphics[width=0.99\textwidth]{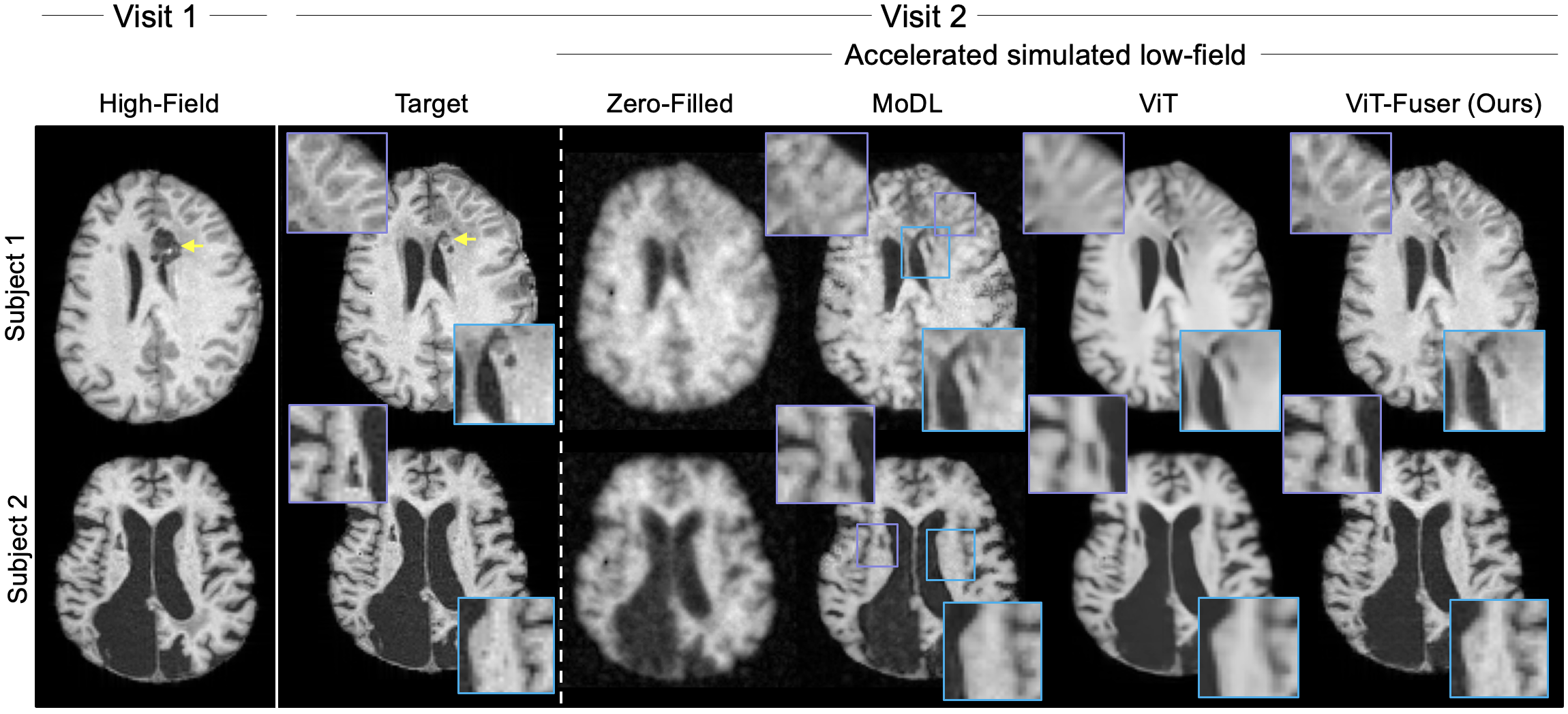}
\caption{\textbf{ViT-Fuser reconstructions compared to other methods: Longitudinal change and similarity preservation experiments.} Comparison of MoDL \cite{Aggarwal2019}, ViT \cite{Lin2022}, and ViT-Fuser (ours) for in-vivo Glioblastoma data \cite{Suter2022} with 8-fold Poisson Disc undersampling with SNR of $10dB$. ViT-Fuser demonstrates decent reconstruction of longitudinal anatomical changes (subject 1, yellow arrow) and temporal similarities (subject 2).
}\label{figPoisson}
\end{figure}

Next, we performed experiments with Poisson Disc \cite{Bridson2007} undersampling with an acceleration factor of $R=8$ and SNR of $10dB$ (Fig. \ref{figPoisson}). As in the previous experiment, ViT-Fuser provides higher reconstruction quality than the competing methods. Notably, the case shown in the top row of this figure exhibits anatomical differences between the baseline and follow-up scans (Fig. \ref{figPoisson}, yellow arrows). This scenario reflects a clinically realistic situation, where changes can occur between repeated visits. Importantly, the results indicate that ViT-Fuser reconstructs the follow-up scan with high accuracy even in the presence of such changes — it does not simply "copy" information from the prior scan, but rather leverages personalized features to enhance reconstruction quality. This robustness to change is essential for longitudinal imaging, where accurate depiction of small changes is critical.

\textbf{\textit{Robustness to SNR variations}}.
Robustness to changes in SNR is particularly critical for reconstruction algorithms designed for low-field systems, as these systems often lack electromagnetic shielding and are highly susceptible to noise. The SNR can also fluctuate significantly due to factors such as system quality, scan duration, and surrounding environmental conditions \cite{poojar2024repeatability}. Therefore, we further examined the robustness of ViT-Fuser to SNR changes.

\begin{figure}[htbp]
\centering
\includegraphics[width=1\textwidth]{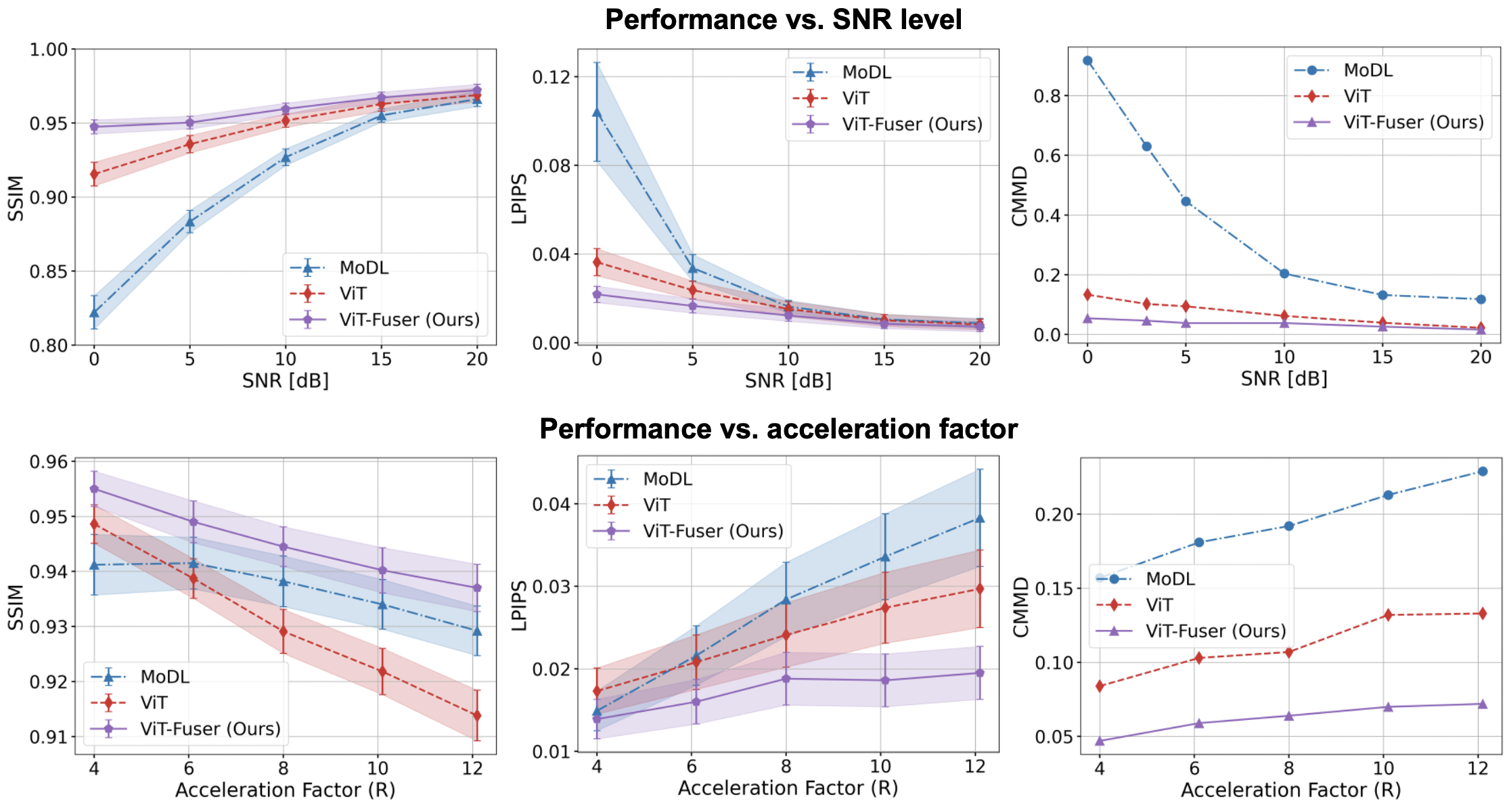}
\caption{\textbf{Robustness to varying SNR levels and acceleration factors.} Top: SSIM, LPIPS, and CMMD metrics versus varying SNR level (dB) for 3-fold equispaced sampling. Bottom: metrics versus acceleration factor using 2D Poisson-disc undersampling at $10dB$ SNR. Note that the SNR axis is in dB, i.e., the shift from $0dB$ to $20dB$ corresponds to a 100-fold linear increase in SNR.
In all cases, our method achieves the best results (highest SSIM, lowest LPIPS and CMMD) and demonstrates high robustness (most moderate slope) across the studied spectrum, i.e. for varying SNR or acceleration factors. 
}\label{metrics_vs_SNR_R}
\end{figure}

Here, we simulated low-field acquisitions for SNR levels of $0dB$ to $20dB$.  The results (Fig. \ref{metrics_vs_SNR_R}, top row) suggest that ViT-Fuser is more robust to such changes than the other two methods (MoDL and ViT). Specifically, it can be seen that ViT-Fuser consistently provides the best results; the image quality metrics are better than those of the other methods (higher SSIM, lower LPIPS and CMMD), their standard deviations are smaller, and the slopes of their curves are the most moderate. All of these indicate high robustness to SNR changes.

\textbf{\textit{Robustness to acceleration factor variations}}.
Next, we examined reconstruction results for acceleration factors of $R=4$ to $R=12$ 
(Fig. \ref{metrics_vs_SNR_R}, bottom row). The results further suggest that our method consistently provides better performance than the competing methods. This is reflected by the values of the quantitative metrics (higher SSIM, lower LPIPS and CMMD). 


In summary, these experiments demonstrate the ability of ViT-Fuser to handle variations in sampling scheme, SNR, and acceleration factor, and to accurately reconstruct anatomical changes.

\subsection{Low-field phantom  data}\label{subsec2.2}


To evaluate our framework using raw low field data acquired under controlled settings, we fabricated two realistic anatomy phantoms \cite{Najac2023balanced} and scanned them in-house. The phantoms had the same underlying structure but exhibited structural differences, to mimic longitudinal scans. A phantom representing the initial state was scanned using a Philips (Best, the Netherlands) high-field ($3T$) system, and a second phantom representing a later state with anatomical differences was scanned with an in-house low-field ($47mT$) MRI system \cite{O’Reilly2020, marques2019low}. Both acquisitions used T1-weighted (T1w) protocols: the high-field scan used a 3D turbo field echo (TFE) sequence, and the low-field scan used a 3D turbo spin echo (TSE) sequence.

The prior high-field image was reconstructed by the Philips system and saved in its standard DICOM format, to represent a realistic scenario where a patient comes with a CD containing an image from a previous scan. The low-field data, on the other hand, were saved as raw k-space data in the numpy format (.npy) , as our method aims to reconstruct images from such raw data.

The pre-trained ViT-Fuser model was fine-tuned using 10 slices chosen randomly from human scans acquired with the same setting (see section \ref{subsec2.3}). The model was then tested on the phantom data without any additional training. To simulate a 4-fold accelerated low-field scan, the data were retrospectively undersampled with equispaced undersampling.



\begin{figure}[htbp]
\centering
\includegraphics[width=1\textwidth]{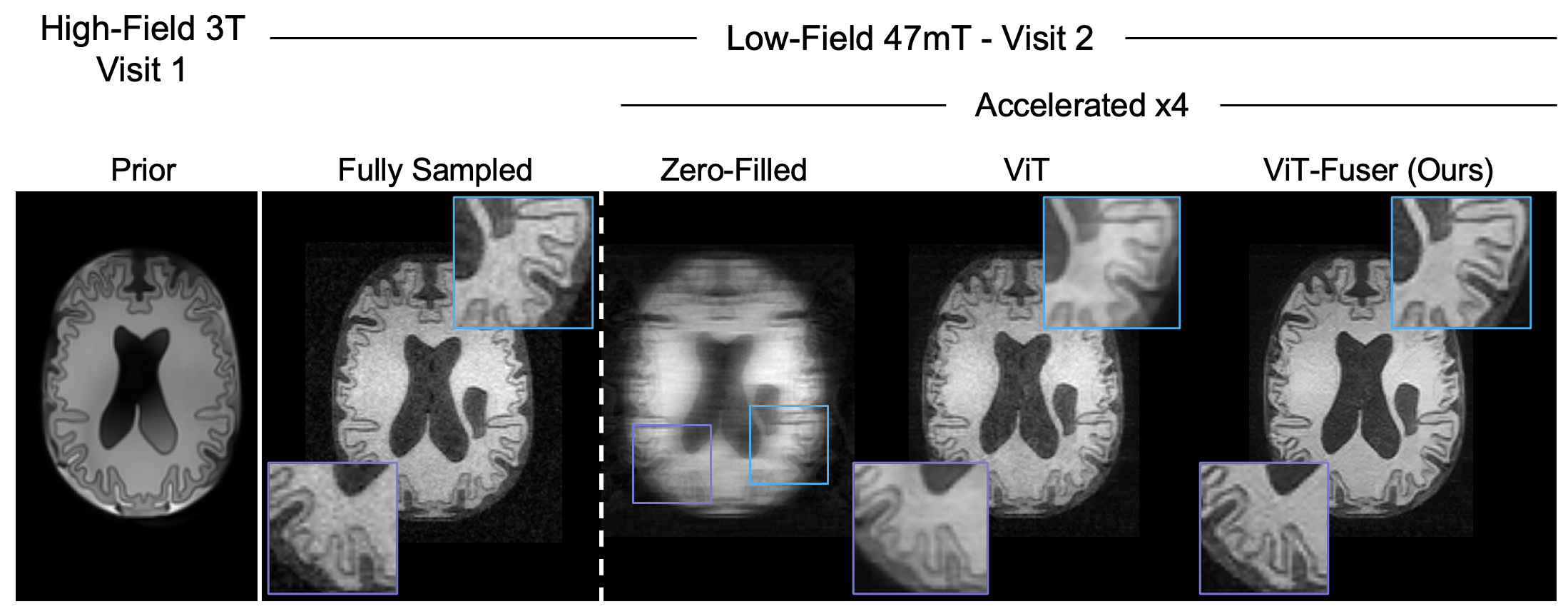}
\caption{\textbf{Demonstration of ViT-Fuser for T1w longitudinal phantoms acquired with $47mT$ scanner paired with a $3T$ T1w prior.}
ViT-Fuser applied to a retrospectively 4-fold equispaced undersampled $47mT$ T1w scan containing a synthetic lesion, paired with a lesion-free high-field $3T$ T1w prior. Compared to the ViT baseline, ViT-Fuser achieves improved image quality.}\label{phantomT1w}
\end{figure}

The results (Fig. \ref{phantomT1w}) demonstrate that ViT-Fuser achieves follow-up reconstructions of higher quality than the ViT model, providing sharper images with less noise and enhanced tissue contrast. Note that this improvement was achieved without sacrificing structural accuracy; ViT-Fuser reconstructed the tumor-mimicking "lesion" accurately, although it did not appear in the prior high-field scan. This further suggests that our approach is capable of harnessing learned features from a prior scan without distorting anatomical changes.

\subsection{Low-field in-vivo data}\label{subsec2.3}

Next, we tested our model with \textit{in-vivo} brain data acquired from two volunteers using the same setting as the previous experiment; the baseline scans were performed with a high-field system (Philips, 3T) and the follow-up scans were performed with an in-house low-field system ($47mT$) \cite{O’Reilly2020, marques2019low}. Here, too, the prior high-field images were reconstructed by the Philips system and saved in a standard DICOM format, and the follow-up data were saved as raw k-space data in the numpy format, to mimic realistic scenarios. Acceleration of the low-field scans was simulated through retrospective 3 and 4-fold equispaced undersampling. As mentioned previously, our ViT-Fuser model was pre-trained on simulated low-field data constructed by degrading high-field data (see section \ref{sec2}). To enable it to adapt to \textit{raw} low-field data, we fine-tuned it using 10 randomly selected slices from one of the volunteers (subject 1). Then, we tested it on data from one additional volunteer and on different slices from that volunteer. 

\begin{figure}[htbp]
\centering
\includegraphics[width=0.94\textwidth]{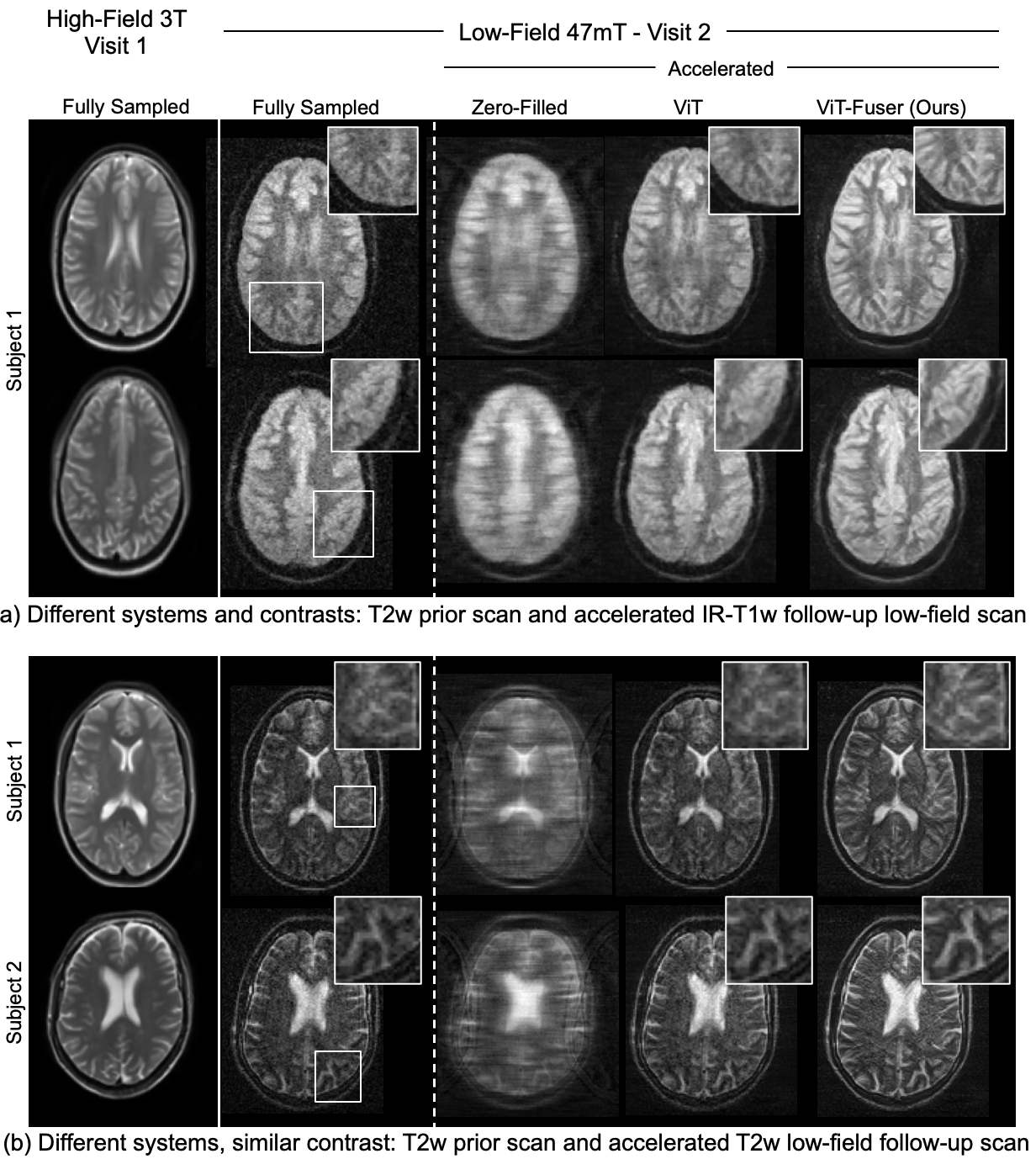}
\caption{\textbf{Demonstration of ViT-Fuser for in-vivo $47mT$ data with a different-contrast $3T$ prior.}
We applied the ViT-Fuser with an accelerated 3-fold equispaced scheme to a $47mT$ Short-IR T1w (a) and 4-fold equispaced scheme to a $47mT$ T2w (b) scans from different subjects alongside a corresponding $3T$ T2w high-field prior. ViT-Fuser exhibits robustness to out-of-distribution priors, enabling improved reconstruction sharpness quality and enhancement of anatomical details.}\label{halbach_invivo}
\end{figure}

Because human tissues exhibit different relaxation parameters in high and low magnetic fields, using the same pulse sequences could provide images with different contrasts \cite{arnold2023low}. In previous work we collaborated with radiologists to find which pulse sequences provide the highest visual similarity between high- and low-field scans \cite{Najac2025repeatability}. Based on that work, here we used the following setting: the prior high-field scans were performed with a T2-weighted pulse sequence and the follow-up scans were performed with a short-inversion-recovery (short-ir) T1-weighted pulse sequence. 



The results (Fig. \ref{halbach_invivo}a) demonstrate that ViT-Fuser provides high-quality low-field images, with less noise and enhanced visibility of anatomical structures compared to the ViT method. Notably, these results were obtained despite the differences in the pulse sequence and tissue contrast between the prior and follow-up scans.

In the second experiment, the prior high-field and follow-up low-field scans were done with T2-weighted pulse sequences (Fig. \ref{halbach_invivo}b). Consistent with previous experiments, the results (Fig. \ref{halbach_invivo}) indicate that our model provides high-quality reconstructions.  This experiment hence further demonstrates that our framework can efficiently bridge the differences between high- and low-field data acquired in different settings. 

\subsection{Case study: out-of-distribution ultra-low-field ($6.5mT$) data}\label{subsec2.4}


We also conducted an experiment with unique in-house data, where the prior scan was done with a clinical $3T$ system (Siemens) and the follow-up scan was done with an ultra-low-field $6.5mT$ system \cite{sarracanie2015low}. Here, the high-field scan was conducted with a T2-weighted pulse sequence and the low-field scan was done with a balanced steady state free precession (bSSFP) sequence, as ultra-low-field scans are highly SNR-starved and bSSFP provides higher SNR than other pulse sequences. As in previous experiments, the prior scan was reconstructed by the vendor's system and saved in a DICOM format, and raw data from the follow-up scan were saved in the numpy format.

Unlike other experiments, here we did not fine-tune the network to the field strength ($6.5mT$) or the pulse sequence (bSSFP) of the raw low-field data, to test our method's ability to generalize to new regimes. Specifically, we used the network that had been pre-trained on the LUMIERE database and then fine-tuned on several slices from the $47mT$ data (see section \ref{subsec5}). We note that despite this fine-tuning, that there was still substantial difference between the training and test settings, i.e. between $47mT$ and the $6.5mT$, because in this low-field regime the SNR varies substantially as factor of the magnetic field strength ($\text{SNR}\propto B_0^{7/4}$). This experiment hence constitutes an out-of-distribution test  for our model.

\begin{figure}[htbp]
\centering
\includegraphics[width=1\textwidth]{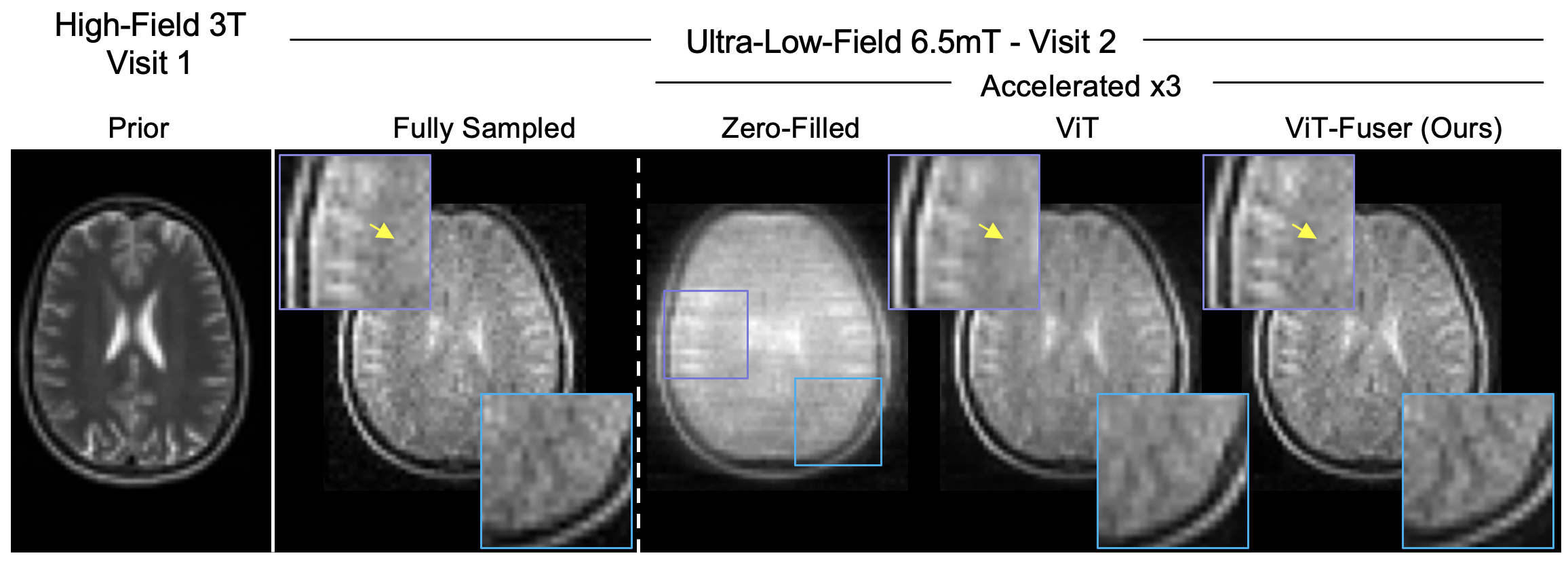}
\caption{\textbf{Demonstration of ViT-Fuser for out-of-distribution contrast and field strength in an in-vivo ultra-low-field scan.} ViT-Fuser applied to an accelerated retrospective 4-fold 2D Variable-Density $6.5mT$ bSSFP scan alongside a corresponding $3T$ T2w high-field prior. ViT-Fuser exhibits robustness to out-of-distribution priors and field strength, enabling improved reconstruction sharpness and contrast.}\label{MattOOD}
\end{figure}

The results (Fig. \ref{MattOOD}), demonstrate that, compared to the ViT method, ViT-Fuser provides cleaner, enhanced images, with higher contrast and better preservation of anatomical structures (see arrow). These findings indicate the framework's ability to generalize to the \textit{ultra-low-field} regime and generalize to acquisition settings that substantially deviate from the training set.


\section{Discussion}\label{sec6}


In this manuscript, we present a novel healthcare paradigm consisting of \textit{multi-field-strength} MRI scans, which leverage the complementary advantages of high- and low-field systems and aims to make longitudinal MRI more accessible and affordable. We also introduced ViT-Fuser, a DL framework that enables the materialization of this strategy. We demonstrated the capabilities of ViT-Fuser via extensive experiments with four datasets, including simulated low-field data based on high-field $1.5T$ scans \cite{Suter2022}, phantom and in-vivo low-field ($47mT$) data, and ultra-low-field ($6.5mT$) in-vivo data; the latter low-field data were paired with prior $3T$ data. These experiments demonstrate the feasibility of our strategy; ViT-Fuser successfully learned features from the prior scans and utilized them to improve image quality and enable acceleration of follow-up low field scans. 

\textit{\textbf{Robustness and generalizability}}.
In our experiments, ViT-Fuser exhibited high robustness to various data types, field strengths, acceleration factors, and tissue contrasts originating from different pulse sequences. Furthermore, the experiment with the $6.5mT$ bSSFP data demonstrates that ViT-Fuser exhibits robustness to out-of-distribution data, as our model was not trained or fine-tuned for this field strength or pulse sequences. Our experiments also showed that ViT-Fuser can generalize to new pathological conditions, e.g.  Glioblastoma tumors, when those appear for the first time in the follow-up low-field scan (Figs. \ref{figPoisson} and \ref{phantomT1w}).

Moreover, ViT-Fuser maintains high performance across a wide range of SNR levels (Fig. \ref{metrics_vs_SNR_R}). This is highly important for low-field systems, as those are especially susceptible to electromagnetic noise variations due to their lack of shielding and mobile design. Such robustness can ensure stable performance in various settings, particularly imaging in point-of-care location. ViT-Fuser is hence a highly robust and general framework, applicable to various clinical settings and commercial systems. 

\textit{\textbf{Relation to other work}}. This work proposes and demonstrates the benefits of using \textit{personalized priors} for improving the performance of low-field MRI scans in clinical workflows. Some previous work considered the use of side-information, e.g. multi-contrast scans or data from past scans, for improving image reconstruction \cite{Weizman2015, Weizman2016, Polak2020, Atalik2024}. However, those studies have only considered settings where the new and previous or "side" data were all acquired in the same field strength. To our knowledge, our work is the first to consider a \textit{multi-field-strength} pipeline that involves low-field MRI and proposes a computational framework to address the challenges that emerge in such pipelines. Those stem from differences in SNR and tissue contrast between high- and low-field scans performed with various systems, as well as anatomical differences that sometimes occur in longitudinal scans. An additional challenge is the lack of online datasets that contain longitudinally paired data, acquired at different time points with high- and low-field scans. To overcome these challenges, we proposed an efficient DL architecture, ViT-Fuser, and a unique training pipeline. Therefore, this work presents novelty with regard to both the conceptual and computational aspects of low-field MRI. 

\textbf{Realistic scenarios: DICOM data from prior scans}. Our experiments with $3T$ prior data and follow-up low-field ($47mT$) data demonstrated the ability of ViT-Fuser to efficiently extract personalized features from images that have been reconstructed by the vendor software and stored in standard DICOM files (Figs. \ref{phantomT1w} and \ref{halbach_invivo}). These experiments hence provide a demonstration of a real-world scenario, when a patient may come with a CD of a previous scan, done with another healthcare provider, and request to use that information in the current scan. We note that in these experiments the DICOM images were fed into the network as input, without any retrospective undersampling or processing, and that \textbf{all} the low-field ($47mT$ or $6.5mT$) data were \textit{raw} data acquired in house. Our approach thus avoided any bias due to off-label data use \cite{Shimron2022} and demonstrated real-world settings.


\textit{\textbf{Limitations}}. This work also has some limitations. Some of our experiments were performed using simulations with a public high-field database \cite{Suter2022}, where images have already been reconstructed and later retrospectively undersampled. However, we have chosen this database due to lack of other suitable open-access data. Furthermore, using a high-field database enabled us to analyze the performance of our method for various SNR settings, as noise was added retrospectively. Moreover, our experiments with \textit{raw} low-field data (see sections \ref{subsec2.2}, \ref{subsec2.3}, and \ref{subsec2.4}) confirmed the predictions from simulations and provided examples with real-world data.

\textit{\textbf{Outlook and applications.}}
The proposed framework was demonstrated for longitudinal brain imaging, where consistent and accurate follow-up is crucial. It has wide clinical applications, including imaging of stroke, neurodegenerative diseases, post-operative monitoring, and cancer surveillance imaging. Moreover, our startegy is not limited to brain imaging, and can also be widely applicable to other settings, including body and cardiac imaging, enabled by recent advances in low-field hardware \cite{zhao2024whole}. Our approach may therefore offer substantial benefits for patients needing longitudinal care, and can be broadly used for various clinical applications, including screening, monitoring, surveillance testing, and treatment response evaluation. This work can hence set the ground for further research on accessible MRI with multi-field-strength pipelines.

\section{Appendix A - Methods (extended)}\label{sec3}
The fundamental MRI reconstruction inverse problem concerns the task of reconstructing an image from sub-Nyquist sampled \kspace{} measurements. A common approach to address this inverse problem is to formulate it as a regularized least-squares optimization:

\begin{equation} \hat{\mathbf{x}} = \arg\min_{\mathbf{x}} || MF\mathbf{x} - \mathbf{y} ||_2^2 + \lambda\cdot R(x)\label{minimization1} \end{equation} where y is the acquired \kspace{} samples, M is an operator that chooses the locations of the sampled \kspace{} pixels, F is the Fourier operator, and $R(x)$ is a regularization term, such as $l_1$ norm in the wavelet domain \cite{Lustig2007}.

Traditional DL frameworks commonly solve the reconstruction problem by training a neural network $f_\theta$ with parameters $\theta$ to map the measurements y to a clean image. The network is typically trained in a supervised manner by minimizing a loss such as:
\begin{equation}
L(\theta) = \frac{1}{N} \sum_{i=1}^{N} \left\| \mathbf{x}^{(i)} - f_{\theta}(\mathbf{Fy}^{(i)}) \right\|_2^2
\label{ClassicLoss} \end{equation} We reformulate the reconstruction problem to include additional prior information from a high-field scan of the same subject. Specifically, we aim to find a reconstruction $\hat{x}$ given the acquired noisy \kspace samples from the low-field accelerated scan and a prior scan $x_{HF}^{prior}$ from high-field scanner. This reformulation introduces a regularization term that enforces consistency between the reconstructed image and the high-field prior as described in Eq. \ref{regularization}:

\begin{equation}
R(x,x_{HF}^{prior}) =  | G\Phi x - G\Phi x_{HF}^{prior} |_1 
\label{regularization} \end{equation} where $\Phi$ is an operator of feature extraction from a pre-trained network, e.g., VGG \cite{Simonyan2014}, and G denotes the operator that calculates the GRAM matrix (dot product between the feature maps and their transposed versions). The combination of feature extraction and GRAM matrix calculation enables capturing an approximation of the feature distribution of both the output and the 'quality' target.

This reformulation of the problem reduces the solution space by incorporating prior knowledge, focusing the search on a smaller, more constrained subset of possible solutions. This reformulation is a minimization problem that facilitates consistency with the measurements while retaining shared information and visually relevant features from the prior scan. 
We propose a novel hybrid loss function that integrates a pixel-space loss with a feature-space loss [24–26].
This new hybrid loss function incorporates our new regularization term $R(x,x_{HF}^{prior})$, which leverages prior knowledge of the target properties of the desired reconstruction properties. Training a DL framework with our approach can follow the loss formulation described in Eq. \(\ref{loss_vivo}\):

\begin{equation}
L =  loss_{\text{pixel space}}(\hat{x},x_{LF}) +\lambda\cdot loss_{\text{feature space}}(\hat{x},x_{HF}^{prior}) \label{loss_vivo}\end{equation} where $loss_{\text{pixel space}}$ represents a pixel-domain loss function (e.g., SSIM, MSE), $loss_{\text{feature space}}$ incorporates our proposed regularization term, and $X_{LF}$ denotes the fully sampled low-field scan.

In simulation settings, as described in the Training Procedure section, the target result is accessible, as $X_{LF} = X_{GT} + n$, allowing a straightforward training loss:

\begin{equation}
L =  loss_{\text{pixel space}}(\hat{x},x_{GT}) +\lambda\cdot loss_{\text{feature space}}(\hat{x},x_{GT}) \label{loss_simulation}\end{equation} where $x_{GT}$ represents the fully sampled, noise-free ground-truth target, and n denotes the additive noise.

It is important to note the distinct contribution of the loss components in the equation. The pixel-space loss, is computed for the reconstructed output and the ground truth low-field image. In contrast, the feature-space loss is computed for the reconstructed output and the ‘quality’ target. Also, the use of feature extraction and the $l_1$ norm (Eq. \ref{regularization}) enables sparse similarity in the feature space, preserves the shared properties of the prior and the current scan and aligns them. This hybrid loss function aims that the reconstructed images will provide both anatomical accuracy, as reflected in high SSIM \cite{Wang2004}, low LPIPS \cite{Zhang2018} and high visual fidelity, as reflected in low CMMD \cite{Jayasumana2023}. 
This trade-off balances low distortion for detail preservation with high perceptual quality to enhance texture differentiation \cite{Blau2018}.

\subsubsection*{Proposed Architecture}\label{Arch_sec}
Transformers have revolutionized machine learning, while Vision Transformer \cite{Dosovitskiy2020} extends their success to computer vision, and recent works also demonstrate decent results for MRI reconstruction \cite{Lin2022, Korkmaz2021, Korkmaz2022, Guo2022}.

We propose ViT-Fuser (Fig. \ref{setup}), a novel transformer-based architecture designed to leverage features from a prior high-field MRI scan of the same subject to enhance the reconstruction of low-field MRI scans.  ViT-Fuser introduces a new feature fusion block that combines extracted features through a learned weighting.
The architecture consists of three main components: (i) dual transformer encoders - two parallel heads, each contains a transformer-based encoder, extract features from the low-field and high-field scans, (ii) feature fusion block - which combines features from both scans and (iii) reconstruction head - a final reconstruction block that reassembles the fused features into the reconstructed and enhanced low-field MRI scan. The feature fusion block uses learned parameters to control the contribution of each scan features, given by the weighted sum of the learnt parameters as in Eq. \eqref{fusion_eq},
\begin{equation}
\phi_{out_k} =\phi_{1_k}\cdot \frac{w_{1_{k}}}{w_{1_{k}}+w_{2_{k}}} + \phi_{2_k}\cdot \frac{w_{2_{k}}}{w_{1_{k}}+w_{2_{k}}} \label{fusion_eq}\end{equation} where $\phi_{i_k}$ denotes the $k$-th feature provided by the transformer encoder output for head $i$,  and w$_{i_k}$ denotes the learnable weight associated with the $k$-th feature of head $i$.  The final reconstruction block reassembles the features into an image.

\subsubsection*{Efficient fusion}\label{subsubsec4}

Our approach provides computational advantages by ensuring that the derivatives of the fusion block remain simple and efficient, enabling rapid back-propagation and stable weight updates. Additionally, its formulation allows simple integration into pre-trained networks, facilitating smooth adaptation without significant modifications.

\textbf{Simple and Efficient Derivatives:}
The derivatives of the fusion block are computationally simple, allowing efficient back-propagation and weight updates. For clarity, we show the derivative of
$\phi_{out_k}$  with respect to \( w_{1_k} \) (noting that the derivative for \( w_{2_k} \) is symmetric), as described in Eq. \ref{eq_derivative1}.

\begin{equation}
\frac{\partial \phi_{out_k}}{\partial w_{1_k}} = (\phi_{1_k} - \phi_{2_k})\cdot \frac{w_{2_k}}{(w_{1_k} + w_{2_k})^2}
\label{eq_derivative1}
\end{equation} Similarly, we calculate the derivative with respect to the input features \( \phi_{1_k} \) to enable efficient gradient propagation back to the transformer encoders (note that the derivative for \( \phi_{2_k} \) is similar), which described in Eq. \ref{eq_derivative_inputs}.

\begin{equation}
\frac{\partial \phi_{out_k}}{\partial \phi_{1_k}} = \frac{w_{1_k}}{w_{1_k} + w_{2_k}}
\label{eq_derivative_inputs}
\end{equation} These derivatives are simple fractions that depend only on the learned weights, providing fast and stable gradient flow to the preceding transformer heads.

\textbf{Training Efficiency:}
This fusion technique is computationally efficient, as it requires only a few arithmetic operations per feature. This simplicity makes it highly scalable for large feature sets and facilitates rapid convergence during training. Moreover, the initialization of weights with intuitive ratios (e.g., 1-0, 0.9-0.1) allows smooth integration with pre-trained vision transformers. 

\subsubsection*{Datasets}\label{subsec3}
The proposed method was trained and tested utilizing various datasets and settings, which demonstrate the effectiveness of our approach for both phantoms and in-vivo data. It also demonstrates success over in-distribution data and even over out-of-distribution data.

\subsubsection*{Longitudinal Glioblastoma data}\label{subsubsec3}

For developing the framework, we utilized the LUMIERE database, which comprises longitudinal scans of Glioblastoma patients \cite{Suter2022}. From this database, we obtained data of 12 subjects, each with two scans acquired at different clinical visits. As about 25 slices were available for each subject, this provided a database of 300 pairs of images. The data were acquired with $3T$ Siemens (Munich) Aera and Avanto systems, with TE of $2.67-2.92ms$ and TR of $1580-1720ms$. All data in the LUMIERE dataset were obtained using high-field MRI systems.

For this study, we structured the dataset to simulate a clinically relevant setting. Specifically, for each subject, the first visit’s scan was retained as high-field, fully sampled data, while the second visit’s scan was processed through a low-field simulation pipeline to emulate real-world constraints. The data were originally stored in DICOM format, hence, to avoid data crimes \cite{Shimron2022}, we provide a detailed description of our full processing pipeline along with demonstrations on in-vivo data.

Both baseline and follow-up scans were given $1.5T$ single coil 3D Inversion-Recovery (IR) with a resolution of 1×1×1$mm$ and matrix size of 256×256. Each two visits were registered by 2D registration with BRISK features \cite{Leutenegger2011} and RANSAC \cite{Fischler1987} based algorithm with MATLAB (Natick, Massachusetts) computer vision toolbox \cite{MathWorksCVT2024}. We have simulated low-field for the second scan of each patient. This simulation included the following steps:
1) resolution degradation to a resolution of $1.5mm$,
2) multiplication of all the data with artificial phase masks \cite{Deveshwar2023}, to generate complex-valued data, and 3) additive white Gaussian noise was applied to the follow-up scan at different SNRs. The baseline scans remained fully sampled, while the follow-up simulated low-field scans were undersampled using different sampling schemes, which are described later.

\subsubsection*{Phantom data: paired high-field ($3T$) and low-field ($47mT$) data}
This data were acquired with $47mT$ Halbach magnet-based MRI system \cite{O’Reilly2020, marques2019low}. We used 3D Turbo Spin Echo (TSE) with TE of $16ms$, TR of $600ms$, resolution of 1.5×1.5×5$mm$ and matrix size of 150×136. The corresponding high-field data were acquired with Philips (Best, the Netherlands) $3T$ T1w Turbo Fast Echo (TFE) with TE of $4.6ms$ and TR of $9.8ms$ scanner at the same resolution.

\subsubsection*{In-vivo dataset: paired high-field ($3T$) and low-field ($47mT$) data}\label{subsubsec4}

To demonstrate the effectivness of our method, we also deployed it on in-vivo T2w 3D- Turbo Spin Echo scans acquired using a low-field $47mT$ Halbach magnet-based MRI system \cite{O’Reilly2020, marques2019low}. These scans were acquired with a $TE = 150ms$, $TR = 2500ms$, a spatial resolution of 1.5×1.5×5mm and a matrix size of 150×136. 
Moreover, we deployed an in-vivo 3D TSE $47mT$ IR-T1w with TE of $20ms$, TR of $1200ms$, TI of $91ms$ and the same resolution.
The corresponding high-field prior scans were acquired using a Philips (Best, the Netherlands) $3T$ T2w 2D multi-slice Fast Spin Echo (FSE) scanner at the same resolution.

\subsubsection*{Out-of-distribution in-vivo data: paired ($3T$) and ultra-low-field ($6.5mT$) data}

Moreover, we demonstrated our approach on an ultra-low-field $6.5mT$ scans, bSSFP sequence with 62 repetitions (NEX) and 69×64 matrix size. For this data we used as prior a corresponding scan from a Siemens (Munich) $3T$ T2w 2D Multi-Slice TSE with a TE of $95ms$, TR of $6100ms$, 0.4×0.4x3$mm$ resolution and 512×512 matrix size.

For all in vivo datasets, the high-field prior was downsampled to match the resolution of the low-field scan and registered using an off-the-shelf method as described earlier.

\subsubsection*{Training procedures}\label{subsec5}
\textbf{Simulation data:}
The training dataset consisted of 220 pairs of longitudinal scans, while the test and validation sets included 50 and 30 pairs, respectively.
We implemented our model with PyTorch \cite{Paszke2019}. To address data limitations, at first, The single-image ViT architecture of Kang and Heckel \cite{Lin2022} was used as the backbone of our ViT-Fuser and was pre-trained on a simple reconstruction task using the LUMIERE database \cite{Suter2022}, augmented with challenging transformations (e.g., extreme rotations and shifts) for 100 epochs. Following this, we extended it to our multi-field-strength architecture, as described above (section \ref{Arch_sec}). This extended architecture was then trained on the same database for 100 epochs using the Adam optimizer \cite{Kingma2014} with SSIM as pixel space loss, after initializing the feature fusion block with $w_1=0.85$ and $w_2=0.15$.

\textbf{In vivo data:} For the in vivo data, initially, the same pre-training procedure as for the simulation data was applied. Finally, the model was fine-tuned on a subset of hard-augmented training slices from the tested $47mT$ scanner with T2w contrast for both visits.

\subsubsection*{Sampling schemes}\label{subsubsec5}
We evaluated our method using two sampling schemes: (i) A one-dimensional equispaced undersampling pattern, implemented using the fastMRI code package \cite{zbontar2018fastMRI} ; (ii) A two-dimensional Poisson Disc undersampling pattern, commonly used for undersampling in three-dimensional acquisitions \cite{Lustig2007}, implemented with the SigPy toolbox \cite{sigpy}.
These undersamplings enabled the examination of the method across both 1D and 2D undersamplings and allowed us to verify the results on both structured (equispaced) and randomized (Poisson-Disc) sampling, utilizing open-source tools.

\subsubsection*{Evaluation metrics}\label{subsec5}
Reconstruction quality was quantitatively assessed using multiple evaluation metrics, including SSIM \cite{Wang2004}, LPIPS \cite{Zhang2018}, and the state-of-the-art CMMD \cite{Jayasumana2023}. These metrics were computed within the region of interest (ROI), defined by a binary mask.
The SSIM metric provides an assessment of visual quality for each reconstruction, while the CMMD metric offers an overall evaluation of the distribution similarity between the target images and the reconstructed images, particularly highlighting the method's ability to preserve tissue structure and texture which becomes increasingly challenging in low SNR regimes.


\section{Disclosures}

\bmhead{Acknowledgements}

T.O. acknowledges support from the Meyer Excellence Award of Technion's Electrical and Computer Engineering Department.
C.F.N and B.L acknowledge funding from Horizon 2020 ERC Advanced PASMAR 101021218 and the Dutch Science Foundation Open Technology 18981. 
A.W. acknowledges support from NWO Open Technology Programme and European Research Council Advanced Grant PASMAR.
E.S. is a Horev Fellow and acknowledges funding support from the Alon Fellowsihp and the Technion's Leaders in Science and Technology program. MSR acknowledges the generous support of the Kiyomi and Ed Baird MGH Research Scholar award. 

We also thank Kang Lin and Reinhard Heckel for their assistance with the ViT architecture and for publicly sharing their code.

\section*{Data availability}
The LUMIERE open dataset can be found online (https://github.com/ysuter/gbm-data-longitudinal).

\section*{Code availability}
The code for this research was implemented in Python and Pytorch was used as the primary package for training the networks. The code will be available online upon publication for an academic purposes.


\section{Appendix B}\label{secA1}

\subsubsection*{Feature-Space Loss Metrics Considerations}
The $L_1$ loss was chosen as the metric applied to the Gram matrix of the extracted features to push towards sparsity. By leveraging $L_1$ loss, we encourage sparse similarity with the target style, enabling the adoption of key properties from the target while preserving the detailed structure of the current scans. This approach avoids excessive averaging or detail suppression, which could arise from using $L_2$ loss. Unlike other alternatives like Frobenius loss, $L_1$ loss does not explicitly weigh the involvement of different features, thereby maintaining a balance between incorporating target style characteristics and retaining the unique details of the current scan.

\subsubsection*{Implementation of other methods}
Our ViT-Fuser was compared to MoDL \cite{Aggarwal2019} and a Vision Transformer \cite{Lin2022}. For the MoDL architecture, we optimized the hyperparameters for four unrolled blocks and eight conjugate gradient (CG) steps in each data consistency block. The MoDL was trained on the mean squared error (MSE) loss, as it provided the best results both perceptually and in terms of metric scores. The Vision Transformer method was implemented using the publicly available code, maintaining the same hyperparameters as in the original work: 10×10 patch size, embedding dimension of 64, depth of 10, and 16 attention heads.
ViT was trained using both MSE loss and our proposed hybrid loss. The hybrid loss yielded superior results both perceptually and in terms of metrics. To further highlight the advantages of our method, we trained ViT with MSE loss to demonstrate the perceptual superiority of our approach, which combines both our architecture and hybrid loss. Additionally, we trained ViT with hybrid loss to showcase the robustness of our architecture across varying SNR levels and acceleration factors.
All architectures were trained using the Adam optimizer \cite{Kingma2014}.

\subsubsection*{Stability for out-of-distribution SNR levels}
\begin{figure}[t]
\centering
\includegraphics[width=0.7\textwidth]{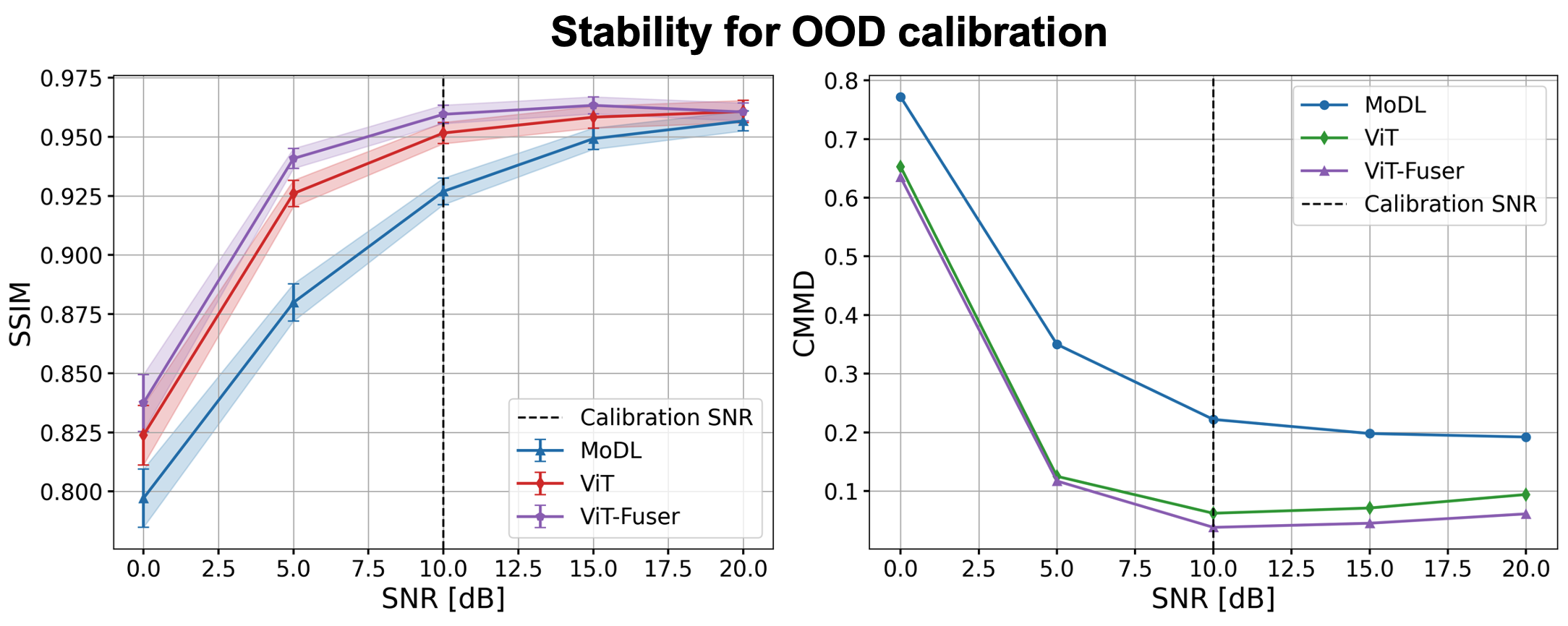}
\caption{\textbf{CMMD and SSIM metric scores as a function of out-of-distribution SNR.} We quantitatively evaluated the reconstruction performance of MoDL, ViT, and ViT-Fuser. The architectures were trained at an SNR of $10dB$ (indicated by the dashed line) with an equispaced 3-fold acceleration. This experiment depicts the stability of ViT-Fuser in reconstruction tasks under varying SNR conditions.
}\label{fig:stability_vs_snr}
\end{figure}

The reconstruction of low-field MRI scans must remain stable across varying SNR levels, as low-field scanner SNR is inherently unstable and varies between scans and different scanners. To evaluate the robustness of our ViT-Fuser, we trained it on an SNR of $10dB$ and tested it across a range of SNR values. The results \ref{fig:stability_vs_snr} demonstrate that ViT-Fuser maintains stability across different SNR conditions, even when the training and inference SNRs differ.

\input{sn-article.bbl}

\end{document}

%% file: sn-article.bbl

%% file: sn-article.bbl
\begin{thebibliography}{69}
\ifx \bisbn   \undefined \def \bisbn  #1{ISBN #1}\fi
\ifx \binits  \undefined \def \binits#1{#1}\fi
\ifx \bauthor  \undefined \def \bauthor#1{#1}\fi
\ifx \batitle  \undefined \def \batitle#1{#1}\fi
\ifx \bjtitle  \undefined \def \bjtitle#1{#1}\fi
\ifx \bvolume  \undefined \def \bvolume#1{\textbf{#1}}\fi
\ifx \byear  \undefined \def \byear#1{#1}\fi
\ifx \bissue  \undefined \def \bissue#1{#1}\fi
\ifx \bfpage  \undefined \def \bfpage#1{#1}\fi
\ifx \blpage  \undefined \def \blpage #1{#1}\fi
\ifx \burl  \undefined \def \burl#1{\textsf{#1}}\fi
\ifx \doiurl  \undefined \def \doiurl#1{\url{https://doi.org/#1}}\fi
\ifx \betal  \undefined \def \betal{\textit{et al.}}\fi
\ifx \binstitute  \undefined \def \binstitute#1{#1}\fi
\ifx \binstitutionaled  \undefined \def \binstitutionaled#1{#1}\fi
\ifx \bctitle  \undefined \def \bctitle#1{#1}\fi
\ifx \beditor  \undefined \def \beditor#1{#1}\fi
\ifx \bpublisher  \undefined \def \bpublisher#1{#1}\fi
\ifx \bbtitle  \undefined \def \bbtitle#1{#1}\fi
\ifx \bedition  \undefined \def \bedition#1{#1}\fi
\ifx \bseriesno  \undefined \def \bseriesno#1{#1}\fi
\ifx \blocation  \undefined \def \blocation#1{#1}\fi
\ifx \bsertitle  \undefined \def \bsertitle#1{#1}\fi
\ifx \bsnm \undefined \def \bsnm#1{#1}\fi
\ifx \bsuffix \undefined \def \bsuffix#1{#1}\fi
\ifx \bparticle \undefined \def \bparticle#1{#1}\fi
\ifx \barticle \undefined \def \barticle#1{#1}\fi
\bibcommenthead
\ifx \bconfdate \undefined \def \bconfdate #1{#1}\fi
\ifx \botherref \undefined \def \botherref #1{#1}\fi
\ifx \url \undefined \def \url#1{\textsf{#1}}\fi
\ifx \bchapter \undefined \def \bchapter#1{#1}\fi
\ifx \bbook \undefined \def \bbook#1{#1}\fi
\ifx \bcomment \undefined \def \bcomment#1{#1}\fi
\ifx \oauthor \undefined \def \oauthor#1{#1}\fi
\ifx \citeauthoryear \undefined \def \citeauthoryear#1{#1}\fi
\ifx \endbibitem  \undefined \def \endbibitem {}\fi
\ifx \bconflocation  \undefined \def \bconflocation#1{#1}\fi
\ifx \arxivurl  \undefined \def \arxivurl#1{\textsf{#1}}\fi
\csname PreBibitemsHook\endcsname

\bibitem[\protect\citeauthoryear{Arnold et~al.}{2023}]{arnold2023low}
\begin{barticle}
\bauthor{\bsnm{Arnold}, \binits{T.C.}},
\bauthor{\bsnm{Freeman}, \binits{C.W.}},
\bauthor{\bsnm{Litt}, \binits{B.}},
\bauthor{\bsnm{Stein}, \binits{J.M.}}:
\batitle{{Low-field MRI: clinical promise and challenges}}.
\bjtitle{Journal of Magnetic Resonance Imaging}
\bvolume{57}(\bissue{1}),
\bfpage{25}--\blpage{44}
(\byear{2023})
\end{barticle}
\endbibitem

\bibitem[\protect\citeauthoryear{Geethanath and Vaughan~Jr}{2019}]{geethanath2019accessible}
\begin{barticle}
\bauthor{\bsnm{Geethanath}, \binits{S.}},
\bauthor{\bsnm{Vaughan~Jr}, \binits{J.T.}}:
\batitle{{Accessible magnetic resonance imaging: A review}}.
\bjtitle{Journal of Magnetic Resonance Imaging}
\bvolume{49}(\bissue{7}),
\bfpage{65}--\blpage{77}
(\byear{2019})
\end{barticle}
\endbibitem

\bibitem[\protect\citeauthoryear{Sarracanie et~al.}{2015}]{sarracanie2015low}
\begin{barticle}
\bauthor{\bsnm{Sarracanie}, \binits{M.}},
\bauthor{\bsnm{LaPierre}, \binits{C.D.}},
\bauthor{\bsnm{Salameh}, \binits{N.}},
\bauthor{\bsnm{Waddington}, \binits{D.E.}},
\bauthor{\bsnm{Witzel}, \binits{T.}},
\bauthor{\bsnm{Rosen}, \binits{M.S.}}:
\batitle{{Low-cost high-performance MRI}}.
\bjtitle{Scientific reports}
\bvolume{5}(\bissue{1}),
\bfpage{15177}
(\byear{2015})
\end{barticle}
\endbibitem

\bibitem[\protect\citeauthoryear{Marques et~al.}{2019}]{marques2019low}
\begin{barticle}
\bauthor{\bsnm{Marques}, \binits{J.P.}},
\bauthor{\bsnm{Simonis}, \binits{F.F.}},
\bauthor{\bsnm{Webb}, \binits{A.G.}}:
\batitle{{Low-field MRI: An MR physics perspective}}.
\bjtitle{Journal of magnetic resonance imaging}
\bvolume{49}(\bissue{6}),
\bfpage{1528}--\blpage{1542}
(\byear{2019})
\end{barticle}
\endbibitem

\bibitem[\protect\citeauthoryear{Cooley et~al.}{2021}]{cooley2021portable}
\begin{barticle}
\bauthor{\bsnm{Cooley}, \binits{C.Z.}},
\bauthor{\bsnm{McDaniel}, \binits{P.C.}},
\bauthor{\bsnm{Stockmann}, \binits{J.P.}},
\bauthor{\bsnm{Srinivas}, \binits{S.A.}},
\bauthor{\bsnm{Cauley}, \binits{S.F.}},
\bauthor{\bsnm{{\'S}liwiak}, \binits{M.}},
\bauthor{\bsnm{Sappo}, \binits{C.R.}},
\bauthor{\bsnm{Vaughn}, \binits{C.F.}},
\bauthor{\bsnm{Guerin}, \binits{B.}},
\bauthor{\bsnm{Rosen}, \binits{M.S.}}, \betal:
\batitle{{A portable scanner for magnetic resonance imaging of the brain}}.
\bjtitle{Nature biomedical engineering}
\bvolume{5}(\bissue{3}),
\bfpage{229}--\blpage{239}
(\byear{2021})
\end{barticle}
\endbibitem

\bibitem[\protect\citeauthoryear{Liu et~al.}{2021}]{liu2021low}
\begin{barticle}
\bauthor{\bsnm{Liu}, \binits{Y.}},
\bauthor{\bsnm{Leong}, \binits{A.T.}},
\bauthor{\bsnm{Zhao}, \binits{Y.}},
\bauthor{\bsnm{Xiao}, \binits{L.}},
\bauthor{\bsnm{Mak}, \binits{H.K.}},
\bauthor{\bsnm{Tsang}, \binits{A.C.O.}},
\bauthor{\bsnm{Lau}, \binits{G.K.}},
\bauthor{\bsnm{Leung}, \binits{G.K.}},
\bauthor{\bsnm{Wu}, \binits{E.X.}}:
\batitle{{A low-cost and shielding-free ultra-low-field brain MRI scanner}}.
\bjtitle{Nature communications}
\bvolume{12}(\bissue{1}),
\bfpage{7238}
(\byear{2021})
\end{barticle}
\endbibitem

\bibitem[\protect\citeauthoryear{Campbell-Washburn et~al.}{2023}]{campbell2023low}
\begin{barticle}
\bauthor{\bsnm{Campbell-Washburn}, \binits{A.E.}},
\bauthor{\bsnm{Keenan}, \binits{K.E.}},
\bauthor{\bsnm{Hu}, \binits{P.}},
\bauthor{\bsnm{Mugler~III}, \binits{J.P.}},
\bauthor{\bsnm{Nayak}, \binits{K.S.}},
\bauthor{\bsnm{Webb}, \binits{A.G.}},
\bauthor{\bsnm{Obungoloch}, \binits{J.}},
\bauthor{\bsnm{Sheth}, \binits{K.N.}},
\bauthor{\bsnm{Hennig}, \binits{J.}},
\bauthor{\bsnm{Rosen}, \binits{M.S.}}, \betal:
\batitle{{Low-field MRI: a report on the 2022 ISMRM workshop}}.
\bjtitle{Magnetic resonance in medicine}
\bvolume{90}(\bissue{4}),
\bfpage{1682}--\blpage{1694}
(\byear{2023})
\end{barticle}
\endbibitem

\bibitem[\protect\citeauthoryear{Kimberly et~al.}{2023}]{Kimberly2023}
\begin{botherref}
\oauthor{\bsnm{Kimberly}, \binits{W.T.}},
\oauthor{\bsnm{Sorby-Adams}, \binits{A.J.}},
\oauthor{\bsnm{Webb}, \binits{A.G.}},
\oauthor{\bsnm{Wu}, \binits{E.X.}},
\oauthor{\bsnm{Beekman}, \binits{R.}},
\oauthor{\bsnm{Bowry}, \binits{R.}},
\oauthor{\bsnm{Schiff}, \binits{S.J.}},
\oauthor{\bsnm{Havenon}, \binits{A.}},
\oauthor{\bsnm{Shen}, \binits{F.X.}},
\oauthor{\bsnm{Sze}, \binits{G.}},
\oauthor{\bsnm{Schaefer}, \binits{P.}},
\oauthor{\bsnm{Iglesias}, \binits{J.E.}},
\oauthor{\bsnm{Rosen}, \binits{M.S.}},
\oauthor{\bsnm{Sheth}, \binits{K.N.}}:
Brain imaging with portable low-field mri.
Nature Reviews Bioengineering
\textbf{1}
(2023)
\doiurl{10.1038/s44222-023-00086-w}
\end{botherref}
\endbibitem

\bibitem[\protect\citeauthoryear{Tian and Nayak}{2024}]{tian2024new}
\begin{barticle}
\bauthor{\bsnm{Tian}, \binits{Y.}},
\bauthor{\bsnm{Nayak}, \binits{K.S.}}:
\batitle{{New clinical opportunities of low-field MRI: heart, lung, body, and musculoskeletal}}.
\bjtitle{Magnetic Resonance Materials in Physics, Biology and Medicine}
\bvolume{37}(\bissue{1}),
\bfpage{1}--\blpage{14}
(\byear{2024})
\end{barticle}
\endbibitem

\bibitem[\protect\citeauthoryear{Obungoloch et~al.}{2023}]{obungoloch2023site}
\begin{barticle}
\bauthor{\bsnm{Obungoloch}, \binits{J.}},
\bauthor{\bsnm{Muhumuza}, \binits{I.}},
\bauthor{\bsnm{Teeuwisse}, \binits{W.}},
\bauthor{\bsnm{Harper}, \binits{J.}},
\bauthor{\bsnm{Etoku}, \binits{I.}},
\bauthor{\bsnm{Asiimwe}, \binits{R.}},
\bauthor{\bsnm{Tusiime}, \binits{P.}},
\bauthor{\bsnm{Gombya}, \binits{G.}},
\bauthor{\bsnm{Mugume}, \binits{C.}},
\bauthor{\bsnm{Namutebi}, \binits{M.H.}}, \betal:
\batitle{{On-site construction of a point-of-care low-field MRI system in Africa}}.
\bjtitle{NMR in Biomedicine}
\bvolume{36}(\bissue{7}),
\bfpage{4917}
(\byear{2023})
\end{barticle}
\endbibitem

\bibitem[\protect\citeauthoryear{Deoni et~al.}{2022}]{deoni2022development}
\begin{barticle}
\bauthor{\bsnm{Deoni}, \binits{S.C.}},
\bauthor{\bsnm{Medeiros}, \binits{P.}},
\bauthor{\bsnm{Deoni}, \binits{A.T.}},
\bauthor{\bsnm{Burton}, \binits{P.}},
\bauthor{\bsnm{Beauchemin}, \binits{J.}},
\bauthor{\bsnm{D’Sa}, \binits{V.}},
\bauthor{\bsnm{Boskamp}, \binits{E.}},
\bauthor{\bsnm{By}, \binits{S.}},
\bauthor{\bsnm{McNulty}, \binits{C.}},
\bauthor{\bsnm{Mileski}, \binits{W.}}, \betal:
\batitle{{Development of a mobile low-field MRI scanner}}.
\bjtitle{Scientific reports}
\bvolume{12}(\bissue{1}),
\bfpage{5690}
(\byear{2022})
\end{barticle}
\endbibitem

\bibitem[\protect\citeauthoryear{Guallart-Naval et~al.}{2022}]{guallart2022portable}
\begin{barticle}
\bauthor{\bsnm{Guallart-Naval}, \binits{T.}},
\bauthor{\bsnm{Algar{\'\i}n}, \binits{J.M.}},
\bauthor{\bsnm{Pellicer-Guridi}, \binits{R.}},
\bauthor{\bsnm{Galve}, \binits{F.}},
\bauthor{\bsnm{Vives-Gilabert}, \binits{Y.}},
\bauthor{\bsnm{Bosch}, \binits{R.}},
\bauthor{\bsnm{Pall{\'a}s}, \binits{E.}},
\bauthor{\bsnm{Gonz{\'a}lez}, \binits{J.M.}},
\bauthor{\bsnm{Rigla}, \binits{J.P.}},
\bauthor{\bsnm{Mart{\'\i}nez}, \binits{P.}}, \betal:
\batitle{{Portable magnetic resonance imaging of patients indoors, outdoors and at home}}.
\bjtitle{Scientific reports}
\bvolume{12}(\bissue{1}),
\bfpage{13147}
(\byear{2022})
\end{barticle}
\endbibitem

\bibitem[\protect\citeauthoryear{Hoult and Richards}{1976}]{hoult1976signal}
\begin{barticle}
\bauthor{\bsnm{Hoult}, \binits{D.I.}},
\bauthor{\bsnm{Richards}, \binits{R.}}:
\batitle{The signal-to-noise ratio of the nuclear magnetic resonance experiment}.
\bjtitle{Journal of Magnetic Resonance (1969)}
\bvolume{24}(\bissue{1}),
\bfpage{71}--\blpage{85}
(\byear{1976})
\end{barticle}
\endbibitem

\bibitem[\protect\citeauthoryear{Shimron et~al.}{2024}]{shimron2024accelerating}
\begin{botherref}
\oauthor{\bsnm{Shimron}, \binits{E.}},
\oauthor{\bsnm{Shan}, \binits{S.}},
\oauthor{\bsnm{Grover}, \binits{J.}},
\oauthor{\bsnm{Koonjoo}, \binits{N.}},
\oauthor{\bsnm{Shen}, \binits{S.}},
\oauthor{\bsnm{Boele}, \binits{T.}},
\oauthor{\bsnm{Sorby-Adams}, \binits{A.J.}},
\oauthor{\bsnm{Kirsch}, \binits{J.E.}},
\oauthor{\bsnm{Rosen}, \binits{M.S.}},
\oauthor{\bsnm{Waddington}, \binits{D.E.}}:
{Accelerating Low-field MRI: Compressed Sensing and AI for fast noise-robust imaging}.
arXiv preprint arXiv:2411.06704
(2024)
\end{botherref}
\endbibitem

\bibitem[\protect\citeauthoryear{Man et~al.}{2023}]{man2023deep}
\begin{barticle}
\bauthor{\bsnm{Man}, \binits{C.}},
\bauthor{\bsnm{Lau}, \binits{V.}},
\bauthor{\bsnm{Su}, \binits{S.}},
\bauthor{\bsnm{Zhao}, \binits{Y.}},
\bauthor{\bsnm{Xiao}, \binits{L.}},
\bauthor{\bsnm{Ding}, \binits{Y.}},
\bauthor{\bsnm{Leung}, \binits{G.K.}},
\bauthor{\bsnm{Leong}, \binits{A.T.}},
\bauthor{\bsnm{Wu}, \binits{E.X.}}:
\batitle{{Deep learning enabled fast 3D brain MRI at 0.055 tesla}}.
\bjtitle{Science Advances}
\bvolume{9}(\bissue{38}),
\bfpage{9327}
(\byear{2023})
\end{barticle}
\endbibitem

\bibitem[\protect\citeauthoryear{Lustig et~al.}{2007}]{Lustig2007}
\begin{botherref}
\oauthor{\bsnm{Lustig}, \binits{M.}},
\oauthor{\bsnm{Donoho}, \binits{D.}},
\oauthor{\bsnm{Pauly}, \binits{J.M.}}:
{Sparse MRI: The application of compressed sensing for rapid MR imaging}.
Magnetic Resonance in Medicine
\textbf{58}
(2007)
\doiurl{10.1002/mrm.21391}
\end{botherref}
\endbibitem

\bibitem[\protect\citeauthoryear{Geethanath et~al.}{2013}]{geethanath2013compressed}
\begin{botherref}
\oauthor{\bsnm{Geethanath}, \binits{S.}},
\oauthor{\bsnm{Reddy}, \binits{R.}},
\oauthor{\bsnm{Konar}, \binits{A.S.}},
\oauthor{\bsnm{Imam}, \binits{S.}},
\oauthor{\bsnm{Sundaresan}, \binits{R.}},
\oauthor{\bsnm{DR}, \binits{R.B.}},
\oauthor{\bsnm{Venkatesan}, \binits{R.}}:
Compressed sensing mri: a review.
Critical Reviews™ in Biomedical Engineering
\textbf{41}(3)
(2013)
\end{botherref}
\endbibitem

\bibitem[\protect\citeauthoryear{Shimron et~al.}{2020}]{shimron2020temporal}
\begin{barticle}
\bauthor{\bsnm{Shimron}, \binits{E.}},
\bauthor{\bsnm{Grissom}, \binits{W.}},
\bauthor{\bsnm{Azhari}, \binits{H.}}:
\batitle{Temporal differences (ted) compressed sensing: A method for fast mrghifu temperature imaging}.
\bjtitle{NMR in Biomedicine}
\bvolume{33}(\bissue{9}),
\bfpage{4352}
(\byear{2020})
\end{barticle}
\endbibitem

\bibitem[\protect\citeauthoryear{Zhu et~al.}{2018}]{zhu2018image}
\begin{barticle}
\bauthor{\bsnm{Zhu}, \binits{B.}},
\bauthor{\bsnm{Liu}, \binits{J.Z.}},
\bauthor{\bsnm{Cauley}, \binits{S.F.}},
\bauthor{\bsnm{Rosen}, \binits{B.R.}},
\bauthor{\bsnm{Rosen}, \binits{M.S.}}:
\batitle{Image reconstruction by domain-transform manifold learning}.
\bjtitle{Nature}
\bvolume{555}(\bissue{7697}),
\bfpage{487}--\blpage{492}
(\byear{2018})
\end{barticle}
\endbibitem

\bibitem[\protect\citeauthoryear{Hammernik et~al.}{2018}]{Hammernik2018}
\begin{botherref}
\oauthor{\bsnm{Hammernik}, \binits{K.}},
\oauthor{\bsnm{Klatzer}, \binits{T.}},
\oauthor{\bsnm{Kobler}, \binits{E.}},
\oauthor{\bsnm{Recht}, \binits{M.P.}},
\oauthor{\bsnm{Sodickson}, \binits{D.K.}},
\oauthor{\bsnm{Pock}, \binits{T.}},
\oauthor{\bsnm{Knoll}, \binits{F.}}:
{Learning a variational network for reconstruction of accelerated MRI data}.
Magnetic Resonance in Medicine
\textbf{79}
(2018)
\doiurl{10.1002/mrm.26977}
\end{botherref}
\endbibitem

\bibitem[\protect\citeauthoryear{Aggarwal et~al.}{2019}]{Aggarwal2019}
\begin{botherref}
\oauthor{\bsnm{Aggarwal}, \binits{H.K.}},
\oauthor{\bsnm{Mani}, \binits{M.P.}},
\oauthor{\bsnm{Jacob}, \binits{M.}}:
{MoDL: Model-Based Deep Learning Architecture for Inverse Problems}.
IEEE Transactions on Medical Imaging
\textbf{38}
(2019)
\doiurl{10.1109/TMI.2018.2865356}
\end{botherref}
\endbibitem

\bibitem[\protect\citeauthoryear{Heckel et~al.}{2024}]{Heckel2024}
\begin{barticle}
\bauthor{\bsnm{Heckel}, \binits{R.}},
\bauthor{\bsnm{Jacob}, \binits{M.}},
\bauthor{\bsnm{Chaudhari}, \binits{A.}},
\bauthor{\bsnm{Perlman}, \binits{O.}},
\bauthor{\bsnm{Shimron}, \binits{E.}}:
\batitle{{Deep learning for accelerated and robust MRI reconstruction}}.
\bjtitle{Magnetic Resonance Materials in Physics, Biology and Medicine 2024 37:3}
\bvolume{37},
\bfpage{335}--\blpage{368}
(\byear{2024})
\doiurl{10.1007/S10334-024-01173-8}
\end{barticle}
\endbibitem

\bibitem[\protect\citeauthoryear{Hammernik et~al.}{2023}]{hammernik2023physics}
\begin{barticle}
\bauthor{\bsnm{Hammernik}, \binits{K.}},
\bauthor{\bsnm{K{\"u}stner}, \binits{T.}},
\bauthor{\bsnm{Yaman}, \binits{B.}},
\bauthor{\bsnm{Huang}, \binits{Z.}},
\bauthor{\bsnm{Rueckert}, \binits{D.}},
\bauthor{\bsnm{Knoll}, \binits{F.}},
\bauthor{\bsnm{Ak{\c{c}}akaya}, \binits{M.}}:
\batitle{Physics-driven deep learning for computational magnetic resonance imaging: Combining physics and machine learning for improved medical imaging}.
\bjtitle{IEEE signal processing magazine}
\bvolume{40}(\bissue{1}),
\bfpage{98}--\blpage{114}
(\byear{2023})
\end{barticle}
\endbibitem

\bibitem[\protect\citeauthoryear{Sharma et~al.}{2022}]{Sharma2022}
\begin{barticle}
\bauthor{\bsnm{Sharma}, \binits{R.}},
\bauthor{\bsnm{Tsiamyrtzis}, \binits{P.}},
\bauthor{\bsnm{Webb}, \binits{A.G.}},
\bauthor{\bsnm{Seimenis}, \binits{I.}},
\bauthor{\bsnm{Loukas}, \binits{C.}},
\bauthor{\bsnm{Leiss}, \binits{E.}},
\bauthor{\bsnm{Tsekos}, \binits{N.V.}}:
\batitle{{A Deep Learning Approach to Upscaling “Low-Quality” MR Images: An In Silico Comparison Study Based on the UNet Framework}}.
\bjtitle{Applied Sciences 2022, Vol. 12, Page 11758}
\bvolume{12},
\bfpage{11758}
(\byear{2022})
\doiurl{10.3390/APP122211758}
\end{barticle}
\endbibitem

\bibitem[\protect\citeauthoryear{Shimron and Perlman}{2023}]{shimron2023ai}
\begin{botherref}
\oauthor{\bsnm{Shimron}, \binits{E.}},
\oauthor{\bsnm{Perlman}, \binits{O.}}:
AI in MRI: computational frameworks for a faster, optimized, and automated imaging workflow.
MDPI
(2023)
\end{botherref}
\endbibitem

\bibitem[\protect\citeauthoryear{Shimron et~al.}{2024}]{shimron2024}
\begin{botherref}
\oauthor{\bsnm{Shimron}, \binits{E.}},
\oauthor{\bsnm{Shan}, \binits{S.}},
\oauthor{\bsnm{Grover}, \binits{J.}},
\oauthor{\bsnm{Koonjoo}, \binits{N.}},
\oauthor{\bsnm{Shen}, \binits{S.}},
\oauthor{\bsnm{Boele}, \binits{T.}},
\oauthor{\bsnm{Sorby-Adams}, \binits{A.J.}},
\oauthor{\bsnm{Kirsch}, \binits{J.E.}},
\oauthor{\bsnm{Rosen}, \binits{M.S.}},
\oauthor{\bsnm{Waddington}, \binits{D.E.J.}}:
{Accelerating Low-field MRI: Compressed Sensing and AI for fast noise-robust imaging}
(2024)
\end{botherref}
\endbibitem

\bibitem[\protect\citeauthoryear{Lau et~al.}{2023}]{Lau2023}
\begin{botherref}
\oauthor{\bsnm{Lau}, \binits{V.}},
\oauthor{\bsnm{Xiao}, \binits{L.}},
\oauthor{\bsnm{Zhao}, \binits{Y.}},
\oauthor{\bsnm{Su}, \binits{S.}},
\oauthor{\bsnm{Ding}, \binits{Y.}},
\oauthor{\bsnm{Man}, \binits{C.}},
\oauthor{\bsnm{Wang}, \binits{X.}},
\oauthor{\bsnm{Tsang}, \binits{A.}},
\oauthor{\bsnm{Cao}, \binits{P.}},
\oauthor{\bsnm{Lau}, \binits{G.K.K.}},
\oauthor{\bsnm{Leung}, \binits{G.K.K.}},
\oauthor{\bsnm{Leong}, \binits{A.T.L.}},
\oauthor{\bsnm{Wu}, \binits{E.X.}}:
{Pushing the limits of low-cost ultra-low-field MRI by dual-acquisition deep learning 3D superresolution}.
Magnetic Resonance in Medicine
\textbf{90}
(2023)
\doiurl{10.1002/mrm.29642}
\end{botherref}
\endbibitem

\bibitem[\protect\citeauthoryear{Koonjoo et~al.}{2021}]{koonjoo2021boosting}
\begin{barticle}
\bauthor{\bsnm{Koonjoo}, \binits{N.}},
\bauthor{\bsnm{Zhu}, \binits{B.}},
\bauthor{\bsnm{Bagnall}, \binits{G.C.}},
\bauthor{\bsnm{Bhutto}, \binits{D.}},
\bauthor{\bsnm{Rosen}, \binits{M.}}:
\batitle{{Boosting the signal-to-noise of low-field MRI with deep learning image reconstruction}}.
\bjtitle{Scientific reports}
\bvolume{11}(\bissue{1}),
\bfpage{8248}
(\byear{2021})
\end{barticle}
\endbibitem

\bibitem[\protect\citeauthoryear{Zhao et~al.}{2024}]{Zhao2024}
\begin{barticle}
\bauthor{\bsnm{Zhao}, \binits{Y.}},
\bauthor{\bsnm{Ding}, \binits{Y.}},
\bauthor{\bsnm{Lau}, \binits{V.}},
\bauthor{\bsnm{Man}, \binits{C.}},
\bauthor{\bsnm{Su}, \binits{S.}},
\bauthor{\bsnm{Xiao}, \binits{L.}},
\bauthor{\bsnm{Leong}, \binits{A.T.L.}},
\bauthor{\bsnm{Wu}, \binits{E.X.}}:
\batitle{{Whole-body magnetic resonance imaging at 0.05 Tesla}}.
\bjtitle{Science (New York, N.Y.)}
\bvolume{384},
\bfpage{7168}
(\byear{2024})
\doiurl{10.1126/SCIENCE.ADM7168/SUPPL_FILE/SCIENCE.ADM7168_MOVIES_S1_TO_S10.ZIP}
\end{barticle}
\endbibitem

\bibitem[\protect\citeauthoryear{Lin et~al.}{2023}]{Lin2023}
\begin{barticle}
\bauthor{\bsnm{Lin}, \binits{H.}},
\bauthor{\bsnm{Figini}, \binits{M.}},
\bauthor{\bsnm{D'Arco}, \binits{F.}},
\bauthor{\bsnm{Ogbole}, \binits{G.}},
\bauthor{\bsnm{Tanno}, \binits{R.}},
\bauthor{\bsnm{Blumberg}, \binits{S.B.}},
\bauthor{\bsnm{Ronan}, \binits{L.}},
\bauthor{\bsnm{Brown}, \binits{B.J.}},
\bauthor{\bsnm{Carmichael}, \binits{D.W.}},
\bauthor{\bsnm{Lagunju}, \binits{I.}},
\bauthor{\bsnm{Cross}, \binits{J.H.}},
\bauthor{\bsnm{Fernandez-Reyes}, \binits{D.}},
\bauthor{\bsnm{Alexander}, \binits{D.C.}}:
\batitle{{Low-field magnetic resonance image enhancement via stochastic image quality transfer}}.
\bjtitle{Medical Image Analysis}
\bvolume{87},
\bfpage{102807}
(\byear{2023})
\doiurl{10.1016/J.MEDIA.2023.102807}
\end{barticle}
\endbibitem

\bibitem[\protect\citeauthoryear{Arefeen et~al.}{2024}]{Arefeen2024}
\begin{botherref}
\oauthor{\bsnm{Arefeen}, \binits{Y.}},
\oauthor{\bsnm{Levac}, \binits{B.}},
\oauthor{\bsnm{Tamir}, \binits{J.I.}}:
{Accelerated, Robust Lower-Field Neonatal MRI with Generative Models}
(2024)
\end{botherref}
\endbibitem

\bibitem[\protect\citeauthoryear{de~Leeuw~den Bouter et~al.}{2022}]{deLeeuwdenBouter2022}
\begin{barticle}
\bauthor{\bsnm{Bouter}, \binits{M.L.}},
\bauthor{\bsnm{Ippolito}, \binits{G.}},
\bauthor{\bsnm{O’Reilly}, \binits{T.P.A.}},
\bauthor{\bsnm{Remis}, \binits{R.F.}},
\bauthor{\bsnm{Gijzen}, \binits{M.B.}},
\bauthor{\bsnm{Webb}, \binits{A.G.}}:
\batitle{{Deep learning-based single image super-resolution for low-field MR brain images}}.
\bjtitle{Scientific Reports}
\bvolume{12},
\bfpage{1}--\blpage{10}
(\byear{2022})
\doiurl{10.1038/S41598-022-10298-6;SUBJMETA=1042,1421,1628,639,692,700,705;KWRD=COMPUTATIONAL+SCIENCE,MAGNETIC+RESONANCE+IMAGING}
\end{barticle}
\endbibitem

\bibitem[\protect\citeauthoryear{Ayde et~al.}{2024}]{Ayde2024}
\begin{barticle}
\bauthor{\bsnm{Ayde}, \binits{R.}},
\bauthor{\bsnm{Vornehm}, \binits{M.}},
\bauthor{\bsnm{Zhao}, \binits{Y.}},
\bauthor{\bsnm{Knoll}, \binits{F.}},
\bauthor{\bsnm{Wu}, \binits{E.X.}},
\bauthor{\bsnm{Sarracanie}, \binits{M.}}:
\batitle{{MRI at low field: A review of software solutions for improving SNR}}.
\bjtitle{Nmr in Biomedicine}
\bvolume{38},
\bfpage{5268}
(\byear{2024})
\doiurl{10.1002/NBM.5268}
\end{barticle}
\endbibitem

\bibitem[\protect\citeauthoryear{Kofler et~al.}{2025}]{kofler2025mr}
\begin{botherref}
\oauthor{\bsnm{Kofler}, \binits{A.}},
\oauthor{\bsnm{Si}, \binits{D.}},
\oauthor{\bsnm{Schote}, \binits{D.}},
\oauthor{\bsnm{Botnar}, \binits{R.M.}},
\oauthor{\bsnm{Kolbitsch}, \binits{C.}},
\oauthor{\bsnm{Prieto}, \binits{C.}}:
Mr imaging in the low-field: Leveraging the power of machine learning.
arXiv preprint arXiv:2501.17211
(2025)
\end{botherref}
\endbibitem

\bibitem[\protect\citeauthoryear{Liu et~al.}{2021}]{Liu2021}
\begin{barticle}
\bauthor{\bsnm{Liu}, \binits{Y.}},
\bauthor{\bsnm{Leong}, \binits{A.T.L.}},
\bauthor{\bsnm{Zhao}, \binits{Y.}},
\bauthor{\bsnm{Xiao}, \binits{L.}},
\bauthor{\bsnm{Mak}, \binits{H.K.F.}},
\bauthor{\bsnm{Tsang}, \binits{A.C.O.}},
\bauthor{\bsnm{Lau}, \binits{G.K.K.}},
\bauthor{\bsnm{Leung}, \binits{G.K.K.}},
\bauthor{\bsnm{Wu}, \binits{E.X.}}:
\batitle{{A low-cost and shielding-free ultra-low-field brain MRI scanner}}.
\bjtitle{Nature Communications 2021 12:1}
\bvolume{12},
\bfpage{1}--\blpage{14}
(\byear{2021})
\doiurl{10.1038/s41467-021-27317-1}
\end{barticle}
\endbibitem

\bibitem[\protect\citeauthoryear{Mazurek et~al.}{2021}]{Mazurek2021}
\begin{barticle}
\bauthor{\bsnm{Mazurek}, \binits{M.H.}},
\bauthor{\bsnm{Cahn}, \binits{B.A.}},
\bauthor{\bsnm{Yuen}, \binits{M.M.}},
\bauthor{\bsnm{Prabhat}, \binits{A.M.}},
\bauthor{\bsnm{Chavva}, \binits{I.R.}},
\bauthor{\bsnm{Shah}, \binits{J.T.}},
\bauthor{\bsnm{Crawford}, \binits{A.L.}},
\bauthor{\bsnm{Welch}, \binits{E.B.}},
\bauthor{\bsnm{Rothberg}, \binits{J.}},
\bauthor{\bsnm{Sacolick}, \binits{L.}},
\bauthor{\bsnm{Poole}, \binits{M.}},
\bauthor{\bsnm{Wira}, \binits{C.}},
\bauthor{\bsnm{Matouk}, \binits{C.C.}},
\bauthor{\bsnm{Ward}, \binits{A.}},
\bauthor{\bsnm{Timario}, \binits{N.}},
\bauthor{\bsnm{Leasure}, \binits{A.}},
\bauthor{\bsnm{Beekman}, \binits{R.}},
\bauthor{\bsnm{Peng}, \binits{T.J.}},
\bauthor{\bsnm{Witsch}, \binits{J.}},
\bauthor{\bsnm{Antonios}, \binits{J.P.}},
\bauthor{\bsnm{Falcone}, \binits{G.J.}},
\bauthor{\bsnm{Gobeske}, \binits{K.T.}},
\bauthor{\bsnm{Petersen}, \binits{N.}},
\bauthor{\bsnm{Schindler}, \binits{J.}},
\bauthor{\bsnm{Sansing}, \binits{L.}},
\bauthor{\bsnm{Gilmore}, \binits{E.J.}},
\bauthor{\bsnm{Hwang}, \binits{D.Y.}},
\bauthor{\bsnm{Kim}, \binits{J.A.}},
\bauthor{\bsnm{Malhotra}, \binits{A.}},
\bauthor{\bsnm{Sze}, \binits{G.}},
\bauthor{\bsnm{Rosen}, \binits{M.S.}},
\bauthor{\bsnm{Kimberly}, \binits{W.T.}},
\bauthor{\bsnm{Sheth}, \binits{K.N.}}:
\batitle{{Portable, bedside, low-field magnetic resonance imaging for evaluation of intracerebral hemorrhage}}.
\bjtitle{Nature Communications}
\bvolume{12},
\bfpage{1}--\blpage{11}
(\byear{2021})
\doiurl{10.1038/S41467-021-25441-6;TECHMETA=57,59;SUBJMETA=308,375,534,575,617,692;KWRD=NEUROLOGY,STROKE,TRANSLATIONAL+RESEARCH}
\end{barticle}
\endbibitem

\bibitem[\protect\citeauthoryear{Sorby-Adams et~al.}{2024}]{Sorby-Adams2024}
\begin{barticle}
\bauthor{\bsnm{Sorby-Adams}, \binits{A.J.}},
\bauthor{\bsnm{Guo}, \binits{J.}},
\bauthor{\bsnm{Laso}, \binits{P.}},
\bauthor{\bsnm{Kirsch}, \binits{J.E.}},
\bauthor{\bsnm{Zabinska}, \binits{J.}},
\bauthor{\bsnm{Guarniz}, \binits{A.L.G.}},
\bauthor{\bsnm{Schaefer}, \binits{P.W.}},
\bauthor{\bsnm{Payabvash}, \binits{S.}},
\bauthor{\bsnm{Havenon}, \binits{A.}},
\bauthor{\bsnm{Rosen}, \binits{M.S.}},
\bauthor{\bsnm{Sheth}, \binits{K.N.}},
\bauthor{\bsnm{Gomez-Isla}, \binits{T.}},
\bauthor{\bsnm{Iglesias}, \binits{J.E.}},
\bauthor{\bsnm{Kimberly}, \binits{W.T.}}:
\batitle{{Portable, low-field magnetic resonance imaging for evaluation of Alzheimer’s disease}}.
\bjtitle{Nature Communications}
\bvolume{15},
\bfpage{1}--\blpage{12}
(\byear{2024})
\doiurl{10.1038/S41467-024-54972-X;TECHMETA=57,59;SUBJMETA=1283,132,375,599,617,692;KWRD=ALZHEIMER}
\end{barticle}
\endbibitem

\bibitem[\protect\citeauthoryear{Shen et~al.}{2024}]{Shen2024}
\begin{barticle}
\bauthor{\bsnm{Shen}, \binits{S.}},
\bauthor{\bsnm{Koonjoo}, \binits{N.}},
\bauthor{\bsnm{Longarino}, \binits{F.K.}},
\bauthor{\bsnm{Lamb}, \binits{L.R.}},
\bauthor{\bsnm{Camacho}, \binits{J.C.V.}},
\bauthor{\bsnm{Hornung}, \binits{T.P.P.}},
\bauthor{\bsnm{Ogier}, \binits{S.E.}},
\bauthor{\bsnm{Yan}, \binits{S.}},
\bauthor{\bsnm{Bortfeld}, \binits{T.R.}},
\bauthor{\bsnm{Saksena}, \binits{M.A.}},
\bauthor{\bsnm{Keenan}, \binits{K.E.}},
\bauthor{\bsnm{Rosen}, \binits{M.S.}}:
\batitle{{Breast imaging with an ultra-low field MRI scanner: a pilot study}}.
\bjtitle{medRxiv : the preprint server for health sciences}
(\byear{2024})
\doiurl{10.1101/2024.04.01.24305081}
\end{barticle}
\endbibitem

\bibitem[\protect\citeauthoryear{Suter et~al.}{2022}]{Suter2022}
\begin{barticle}
\bauthor{\bsnm{Suter}, \binits{Y.}},
\bauthor{\bsnm{Knecht}, \binits{U.}},
\bauthor{\bsnm{Valenzuela}, \binits{W.}},
\bauthor{\bsnm{Notter}, \binits{M.}},
\bauthor{\bsnm{Hewer}, \binits{E.}},
\bauthor{\bsnm{Schucht}, \binits{P.}},
\bauthor{\bsnm{Wiest}, \binits{R.}},
\bauthor{\bsnm{Reyes}, \binits{M.}}:
\batitle{{The LUMIERE dataset: Longitudinal Glioblastoma MRI with expert RANO evaluation}}.
\bjtitle{Scientific Data 2022 9:1}
\bvolume{9},
\bfpage{1}--\blpage{8}
(\byear{2022})
\doiurl{10.1038/s41597-022-01881-7}
\end{barticle}
\endbibitem

\bibitem[\protect\citeauthoryear{Oved and Shimron}{2025}]{Oved2025}
\begin{bchapter}
\bauthor{\bsnm{Oved}, \binits{T.}},
\bauthor{\bsnm{Shimron}, \binits{E.}}:
\bctitle{{Nex-Gen Personalized MRI: Boosting Low-Field MRI Reconstruction with a Feature-Fusion Transformer and High-Field Priors}}.
In: \bbtitle{Proceedings of the International Society of Magnetic Resonance in Medicine (ISMRM) Annual Meeting}
(\byear{2025}).
\bcomment{Abstract \#0343}
\end{bchapter}
\endbibitem

\bibitem[\protect\citeauthoryear{Leutenegger et~al.}{2011}]{Leutenegger2011}
\begin{botherref}
\oauthor{\bsnm{Leutenegger}, \binits{S.}},
\oauthor{\bsnm{Chli}, \binits{M.}},
\oauthor{\bsnm{Siegwart}, \binits{R.Y.}}:
{BRISK: Binary Robust invariant scalable keypoints}.
Proceedings of the IEEE International Conference on Computer Vision,
2548--2555
(2011)
\doiurl{10.1109/ICCV.2011.6126542}
\end{botherref}
\endbibitem

\bibitem[\protect\citeauthoryear{Fischler and Bolles}{1987}]{Fischler1987}
\begin{botherref}
\oauthor{\bsnm{Fischler}, \binits{M.A.}},
\oauthor{\bsnm{Bolles}, \binits{R.C.}}:
{Random Sample Consensus: A Paradigm for Model Fitting with Applications to Image Analysis and Automated Cartography}.
Readings in Computer Vision,
726--740
(1987)
\doiurl{10.1016/B978-0-08-051581-6.50070-2}
\end{botherref}
\endbibitem

\bibitem[\protect\citeauthoryear{{The MathWorks, Inc.}}{2022}]{MathWorksCVT2024}
\begin{bbook}
\bauthor{\bsnm{{The MathWorks, Inc.}}}:
\bbtitle{{Computer Vision Toolbox}}.
\bpublisher{The MathWorks, Inc.},
\blocation{Natick, Massachusetts, USA}
(\byear{2022}).
\bcomment{The MathWorks, Inc.. \url{https://www.mathworks.com/products/computer-vision.html}}
\end{bbook}
\endbibitem

\bibitem[\protect\citeauthoryear{Shimron et~al.}{2022}]{Shimron2022}
\begin{barticle}
\bauthor{\bsnm{Shimron}, \binits{E.}},
\bauthor{\bsnm{Tamir}, \binits{J.I.}},
\bauthor{\bsnm{Wang}, \binits{K.}},
\bauthor{\bsnm{Lustig}, \binits{M.}}:
\batitle{{Implicit data crimes: Machine learning bias arising from misuse of public data}}.
\bjtitle{Proceedings of the National Academy of Sciences of the United States of America}
\bvolume{119},
\bfpage{2117203119}
(\byear{2022})
\doiurl{10.1073/PNAS.2117203119/SUPPL_FILE/PNAS.2117203119.SAPP.PDF}
\end{barticle}
\endbibitem

\bibitem[\protect\citeauthoryear{Lin and Heckel}{2022}]{Lin2022}
\begin{bchapter}
\bauthor{\bsnm{Lin}, \binits{K.}},
\bauthor{\bsnm{Heckel}, \binits{R.}}:
\bctitle{{Vision Transformers Enable Fast and Robust Accelerated MRI}}.
In: \bbtitle{Proceedings of Machine Learning Research},
vol. \bseriesno{172}
(\byear{2022})
\end{bchapter}
\endbibitem

\bibitem[\protect\citeauthoryear{Wang et~al.}{2004}]{Wang2004}
\begin{botherref}
\oauthor{\bsnm{Wang}, \binits{Z.}},
\oauthor{\bsnm{Bovik}, \binits{A.C.}},
\oauthor{\bsnm{Sheikh}, \binits{H.R.}},
\oauthor{\bsnm{Simoncelli}, \binits{E.P.}}:
Image quality assessment: From error visibility to structural similarity.
IEEE Transactions on Image Processing
\textbf{13}
(2004)
\doiurl{10.1109/TIP.2003.819861}
\end{botherref}
\endbibitem

\bibitem[\protect\citeauthoryear{Zhang et~al.}{2018}]{Zhang2018}
\begin{botherref}
\oauthor{\bsnm{Zhang}, \binits{R.}},
\oauthor{\bsnm{Isola}, \binits{P.}},
\oauthor{\bsnm{Efros}, \binits{A.A.}},
\oauthor{\bsnm{Shechtman}, \binits{E.}},
\oauthor{\bsnm{Wang}, \binits{O.}}:
{The Unreasonable Effectiveness of Deep Features as a Perceptual Metric}.
Proceedings of the IEEE Computer Society Conference on Computer Vision and Pattern Recognition,
586--595
(2018)
\doiurl{10.1109/CVPR.2018.00068}
\end{botherref}
\endbibitem

\bibitem[\protect\citeauthoryear{Jayasumana et~al.}{2023}]{Jayasumana2023}
\begin{botherref}
\oauthor{\bsnm{Jayasumana}, \binits{S.}},
\oauthor{\bsnm{Ramalingam}, \binits{S.}},
\oauthor{\bsnm{Veit}, \binits{A.}},
\oauthor{\bsnm{Glasner}, \binits{D.}},
\oauthor{\bsnm{Chakrabarti}, \binits{A.}},
\oauthor{\bsnm{Kumar}, \binits{S.}},
\oauthor{\bsnm{Research}, \binits{G.}},
\oauthor{\bsnm{York}, \binits{N.}}:
{Rethinking FID: Towards a Better Evaluation Metric for Image Generation}
(2023)
\end{botherref}
\endbibitem

\bibitem[\protect\citeauthoryear{Bridson}{2007}]{Bridson2007}
\begin{barticle}
\bauthor{\bsnm{Bridson}, \binits{R.}}:
\batitle{{Fast poisson disk sampling in arbitrary dimensions}}.
\bjtitle{ACM SIGGRAPH 2007 Sketches, SIGGRAPH'07}
(\byear{2007})
\doiurl{10.1145/1278780.1278807/SUPPL_FILE/A22-BRIDSON.ZIP}
\end{barticle}
\endbibitem

\bibitem[\protect\citeauthoryear{Poojar et~al.}{2024}]{poojar2024repeatability}
\begin{botherref}
\oauthor{\bsnm{Poojar}, \binits{P.}},
\oauthor{\bsnm{Oiye}, \binits{I.E.}},
\oauthor{\bsnm{Aggarwal}, \binits{K.}},
\oauthor{\bsnm{Jimeno}, \binits{M.M.}},
\oauthor{\bsnm{Vaughan}, \binits{J.T.}},
\oauthor{\bsnm{Geethanath}, \binits{S.}}:
{Repeatability of image quality in very low-field MRI}.
NMR in Biomedicine,
5198
(2024)
\end{botherref}
\endbibitem

\bibitem[\protect\citeauthoryear{Najac et~al.}{2023}]{Najac2023balanced}
\begin{bchapter}
\bauthor{\bsnm{Najac}, \binits{C.}},
\bauthor{\bsnm{Birk}, \binits{F.}},
\bauthor{\bsnm{O’Reilly}, \binits{T.}},
\bauthor{\bsnm{Scheffler}, \binits{K.}},
\bauthor{\bsnm{Webb}, \binits{A.}},
\bauthor{\bsnm{Heule}, \binits{R.}}:
\bctitle{{Balanced steady-state free precession imaging and associated rapid relaxation time mapping on a point-of-care 46 mT Halbach MRI scanner}}.
In: \bbtitle{Proceedings of the International Society of Magnetic Resonance in Medicine (ISMRM) Annual Meeting},
\bconflocation{Toronto, Canada}
(\byear{2023}).
\bcomment{International Society for Magnetic Resonance in Medicine. Abstract \#1587}
\end{bchapter}
\endbibitem

\bibitem[\protect\citeauthoryear{O’Reilly et~al.}{2020}]{O’Reilly2020}
\begin{barticle}
\bauthor{\bsnm{O’Reilly}, \binits{T.}},
\bauthor{\bsnm{Teeuwisse}, \binits{W.M.}},
\bauthor{\bsnm{Gans}, \binits{D.}},
\bauthor{\bsnm{Koolstra}, \binits{K.}},
\bauthor{\bsnm{Webb}, \binits{A.G.}}:
\batitle{{In vivo 3D brain and extremity MRI at 50 mT using a permanent magnet Halbach array}}.
\bjtitle{Magnetic Resonance in Medicine}
\bvolume{85},
\bfpage{495}
(\byear{2020})
\doiurl{10.1002/MRM.28396}
\end{barticle}
\endbibitem

\bibitem[\protect\citeauthoryear{Najac et~al.}{2025}]{Najac2025repeatability}
\begin{bchapter}
\bauthor{\bsnm{Najac}, \binits{C.}},
\bauthor{\bsnm{Broek}, \binits{R.}},
\bauthor{\bsnm{O'Reilly}, \binits{T.}},
\bauthor{\bsnm{Webb}, \binits{A.}},
\bauthor{\bsnm{Lena}, \binits{B.}}:
\bctitle{{Evaluating repeatability of in vivo imaging in multiple locations using a portable Halbach-based 46 mT scanner}}.
In: \bbtitle{Proceedings of the International Society of Magnetic Resonance in Medicine (ISMRM) Annual Meeting},
\bconflocation{Honolulu, Hawaii, USA}
(\byear{2025}).
\bcomment{International Society for Magnetic Resonance in Medicine. Abstract \#0489}
\end{bchapter}
\endbibitem

\bibitem[\protect\citeauthoryear{Weizman et~al.}{2015}]{Weizman2015}
\begin{barticle}
\bauthor{\bsnm{Weizman}, \binits{L.}},
\bauthor{\bsnm{Eldar}, \binits{Y.C.}},
\bauthor{\bsnm{Bashat}, \binits{D.B.}}:
\batitle{{Compressed sensing for longitudinal MRI: An adaptive-weighted approach}}.
\bjtitle{Medical Physics}
\bvolume{42},
\bfpage{5195}--\blpage{5208}
(\byear{2015})
\doiurl{10.1118/1.4928148;JOURNAL:JOURNAL:24734209;REQUESTEDJOURNAL:JOURNAL:24734209;WGROUP:STRING:PUBLICATION}
\end{barticle}
\endbibitem

\bibitem[\protect\citeauthoryear{Weizman et~al.}{2016}]{Weizman2016}
\begin{barticle}
\bauthor{\bsnm{Weizman}, \binits{L.}},
\bauthor{\bsnm{Eldar}, \binits{Y.C.}},
\bauthor{\bsnm{Bashat}, \binits{D.B.}}:
\batitle{{Reference-based MRI}}.
\bjtitle{Medical Physics}
\bvolume{43},
\bfpage{5357}--\blpage{5369}
(\byear{2016})
\doiurl{10.1118/1.4962032,}
\end{barticle}
\endbibitem

\bibitem[\protect\citeauthoryear{Polak et~al.}{2020}]{Polak2020}
\begin{barticle}
\bauthor{\bsnm{Polak}, \binits{D.}},
\bauthor{\bsnm{Cauley}, \binits{S.}},
\bauthor{\bsnm{Bilgic}, \binits{B.}},
\bauthor{\bsnm{Gong}, \binits{E.}},
\bauthor{\bsnm{Bachert}, \binits{P.}},
\bauthor{\bsnm{Adalsteinsson}, \binits{E.}},
\bauthor{\bsnm{Setsompop}, \binits{K.}}:
\batitle{{Joint multi-contrast variational network reconstruction (jVN) with application to rapid 2D and 3D imaging}}.
\bjtitle{Magnetic Resonance in Medicine}
\bvolume{84},
\bfpage{1456}--\blpage{1469}
(\byear{2020})
\doiurl{10.1002/MRM.28219,}
\end{barticle}
\endbibitem

\bibitem[\protect\citeauthoryear{Atalik et~al.}{}]{Atalik2024}
\begin{botherref}
\oauthor{\bsnm{Atalik}, \binits{A.}},
\oauthor{\bsnm{Chopra}, \binits{S.}},
\oauthor{\bsnm{Sodickson}, \binits{D.}}:
{Leveraging Side Information with Deep Learning for Linear Inverse Problems: Applications to MR Image Reconstruction}
\end{botherref}
\endbibitem

\bibitem[\protect\citeauthoryear{Zhao et~al.}{2024}]{zhao2024whole}
\begin{barticle}
\bauthor{\bsnm{Zhao}, \binits{Y.}},
\bauthor{\bsnm{Ding}, \binits{Y.}},
\bauthor{\bsnm{Lau}, \binits{V.}},
\bauthor{\bsnm{Man}, \binits{C.}},
\bauthor{\bsnm{Su}, \binits{S.}},
\bauthor{\bsnm{Xiao}, \binits{L.}},
\bauthor{\bsnm{Leong}, \binits{A.T.}},
\bauthor{\bsnm{Wu}, \binits{E.X.}}:
\batitle{Whole-body magnetic resonance imaging at 0.05 tesla}.
\bjtitle{Science}
\bvolume{384}(\bissue{6696}),
\bfpage{7168}
(\byear{2024})
\end{barticle}
\endbibitem

\bibitem[\protect\citeauthoryear{Simonyan and Zisserman}{2014}]{Simonyan2014}
\begin{botherref}
\oauthor{\bsnm{Simonyan}, \binits{K.}},
\oauthor{\bsnm{Zisserman}, \binits{A.}}:
{Very Deep Convolutional Networks for Large-Scale Image Recognition}.
3rd International Conference on Learning Representations, ICLR 2015 - Conference Track Proceedings
(2014)
\end{botherref}
\endbibitem

\bibitem[\protect\citeauthoryear{Blau and Michaeli}{2018}]{Blau2018}
\begin{bchapter}
\bauthor{\bsnm{Blau}, \binits{Y.}},
\bauthor{\bsnm{Michaeli}, \binits{T.}}:
\bctitle{{The Perception-Distortion Tradeoff}}.
In: \bbtitle{Proceedings of the IEEE Computer Society Conference on Computer Vision and Pattern Recognition}
(\byear{2018}).
\doiurl{10.1109/CVPR.2018.00652}
\end{bchapter}
\endbibitem

\bibitem[\protect\citeauthoryear{Dosovitskiy et~al.}{2020}]{Dosovitskiy2020}
\begin{botherref}
\oauthor{\bsnm{Dosovitskiy}, \binits{A.}},
\oauthor{\bsnm{Beyer}, \binits{L.}},
\oauthor{\bsnm{Kolesnikov}, \binits{A.}},
\oauthor{\bsnm{Weissenborn}, \binits{D.}},
\oauthor{\bsnm{Zhai}, \binits{X.}},
\oauthor{\bsnm{Unterthiner}, \binits{T.}},
\oauthor{\bsnm{Dehghani}, \binits{M.}},
\oauthor{\bsnm{Minderer}, \binits{M.}},
\oauthor{\bsnm{Heigold}, \binits{G.}},
\oauthor{\bsnm{Gelly}, \binits{S.}},
\oauthor{\bsnm{Uszkoreit}, \binits{J.}},
\oauthor{\bsnm{Houlsby}, \binits{N.}}:
{An Image is Worth 16x16 Words: Transformers for Image Recognition at Scale}.
ICLR 2021 - 9th International Conference on Learning Representations
(2020)
\end{botherref}
\endbibitem

\bibitem[\protect\citeauthoryear{Korkmaz et~al.}{2021}]{Korkmaz2021}
\begin{barticle}
\bauthor{\bsnm{Korkmaz}, \binits{Y.}},
\bauthor{\bsnm{Yurt}, \binits{M.}},
\bauthor{\bsnm{Dar}, \binits{S.U.H.}},
\bauthor{\bsnm{Özbey}, \binits{M.}},
\bauthor{\bsnm{Cukur}, \binits{T.}}:
\batitle{{Deep MRI Reconstruction with Generative Vision Transformers}}.
\bjtitle{Lecture Notes in Computer Science (including subseries Lecture Notes in Artificial Intelligence and Lecture Notes in Bioinformatics)}
\bvolume{12964 LNCS},
\bfpage{54}--\blpage{64}
(\byear{2021})
\doiurl{10.1007/978-3-030-88552-6_6/TABLES/1}
\end{barticle}
\endbibitem

\bibitem[\protect\citeauthoryear{Korkmaz et~al.}{2022}]{Korkmaz2022}
\begin{botherref}
\oauthor{\bsnm{Korkmaz}, \binits{Y.}},
\oauthor{\bsnm{Dar}, \binits{S.U.H.}},
\oauthor{\bsnm{Yurt}, \binits{M.}},
\oauthor{\bsnm{Ozbey}, \binits{M.}},
\oauthor{\bsnm{Cukur}, \binits{T.}}:
{Unsupervised MRI Reconstruction via Zero-Shot Learned Adversarial Transformers}.
IEEE Transactions on Medical Imaging
\textbf{41}
(2022)
\doiurl{10.1109/TMI.2022.3147426}
\end{botherref}
\endbibitem

\bibitem[\protect\citeauthoryear{Guo et~al.}{2022}]{Guo2022}
\begin{barticle}
\bauthor{\bsnm{Guo}, \binits{P.}},
\bauthor{\bsnm{Mei}, \binits{Y.}},
\bauthor{\bsnm{Zhou}, \binits{J.}},
\bauthor{\bsnm{Jiang}, \binits{S.}},
\bauthor{\bsnm{Patel}, \binits{V.M.}}:
\batitle{{ReconFormer: Accelerated MRI Reconstruction Using Recurrent Transformer}}.
\bjtitle{IEEE Transactions on Medical Imaging}
\bvolume{43},
\bfpage{582}--\blpage{593}
(\byear{2022})
\doiurl{10.1109/TMI.2023.3314747}
\end{barticle}
\endbibitem

\bibitem[\protect\citeauthoryear{Deveshwar et~al.}{2023}]{Deveshwar2023}
\begin{botherref}
\oauthor{\bsnm{Deveshwar}, \binits{N.}},
\oauthor{\bsnm{Rajagopal}, \binits{A.}},
\oauthor{\bsnm{Sahin}, \binits{S.}},
\oauthor{\bsnm{Shimron}, \binits{E.}},
\oauthor{\bsnm{Larson}, \binits{P.E.Z.}}:
{Synthesizing Complex-Valued Multicoil MRI Data from Magnitude-Only Images}.
Bioengineering
\textbf{10}
(2023)
\doiurl{10.3390/bioengineering10030358}
\end{botherref}
\endbibitem

\bibitem[\protect\citeauthoryear{Paszke et~al.}{2019}]{Paszke2019}
\begin{botherref}
\oauthor{\bsnm{Paszke}, \binits{A.}},
\oauthor{\bsnm{Gross}, \binits{S.}},
\oauthor{\bsnm{Massa}, \binits{F.}},
\oauthor{\bsnm{Lerer}, \binits{A.}},
\oauthor{\bsnm{Bradbury}, \binits{J.}},
\oauthor{\bsnm{Chanan}, \binits{G.}},
\oauthor{\bsnm{Killeen}, \binits{T.}},
\oauthor{\bsnm{Lin}, \binits{Z.}},
\oauthor{\bsnm{Gimelshein}, \binits{N.}},
\oauthor{\bsnm{Antiga}, \binits{L.}},
\oauthor{\bsnm{Desmaison}, \binits{A.}},
\oauthor{\bsnm{Köpf}, \binits{A.}},
\oauthor{\bsnm{Yang}, \binits{E.}},
\oauthor{\bsnm{DeVito}, \binits{Z.}},
\oauthor{\bsnm{Raison}, \binits{M.}},
\oauthor{\bsnm{Tejani}, \binits{A.}},
\oauthor{\bsnm{Chilamkurthy}, \binits{S.}},
\oauthor{\bsnm{Steiner}, \binits{B.}},
\oauthor{\bsnm{Fang}, \binits{L.}},
\oauthor{\bsnm{Bai}, \binits{J.}},
\oauthor{\bsnm{Chintala}, \binits{S.}}:
{PyTorch: An Imperative Style, High-Performance Deep Learning Library}.
Advances in Neural Information Processing Systems
\textbf{32}
(2019)
\end{botherref}
\endbibitem

\bibitem[\protect\citeauthoryear{Kingma and Ba}{2014}]{Kingma2014}
\begin{botherref}
\oauthor{\bsnm{Kingma}, \binits{D.P.}},
\oauthor{\bsnm{Ba}, \binits{J.L.}}:
{Adam: A Method for Stochastic Optimization}.
3rd International Conference on Learning Representations, ICLR 2015 - Conference Track Proceedings
(2014)
\end{botherref}
\endbibitem

\bibitem[\protect\citeauthoryear{Zbontar et~al.}{2018}]{zbontar2018fastMRI}
\begin{botherref}
\oauthor{\bsnm{Zbontar}, \binits{J.}},
\oauthor{\bsnm{Knoll}, \binits{F.}},
\oauthor{\bsnm{Sriram}, \binits{A.}},
\oauthor{\bsnm{Murrell}, \binits{T.}},
\oauthor{\bsnm{Huang}, \binits{Z.}},
\oauthor{\bsnm{Muckley}, \binits{M.J.}},
\oauthor{\bsnm{Defazio}, \binits{A.}},
\oauthor{\bsnm{Stern}, \binits{R.}},
\oauthor{\bsnm{Johnson}, \binits{P.}},
\oauthor{\bsnm{Bruno}, \binits{M.}},
\oauthor{\bsnm{Parente}, \binits{M.}},
\oauthor{\bsnm{Geras}, \binits{K.J.}},
\oauthor{\bsnm{Katsnelson}, \binits{J.}},
\oauthor{\bsnm{Chandarana}, \binits{H.}},
\oauthor{\bsnm{Zhang}, \binits{Z.}},
\oauthor{\bsnm{Drozdzal}, \binits{M.}},
\oauthor{\bsnm{Romero}, \binits{A.}},
\oauthor{\bsnm{Rabbat}, \binits{M.}},
\oauthor{\bsnm{Vincent}, \binits{P.}},
\oauthor{\bsnm{Yakubova}, \binits{N.}},
\oauthor{\bsnm{Pinkerton}, \binits{J.}},
\oauthor{\bsnm{Wang}, \binits{D.}},
\oauthor{\bsnm{Owens}, \binits{E.}},
\oauthor{\bsnm{Zitnick}, \binits{C.L.}},
\oauthor{\bsnm{Recht}, \binits{M.P.}},
\oauthor{\bsnm{Sodickson}, \binits{D.K.}},
\oauthor{\bsnm{Lui}, \binits{Y.W.}}:
{fastMRI}: An Open Dataset and Benchmarks for Accelerated {MRI}
(2018)
\end{botherref}
\endbibitem

\bibitem[\protect\citeauthoryear{Frank~Ong}{2019}]{sigpy}
\begin{bchapter}
\bauthor{\bsnm{Frank~Ong}, \binits{M.L.}}:
\bctitle{{SigPy: A Python Package for High Performance Iterative Reconstruction}}.
In: \bbtitle{Proceedings of the ISMRM 27th Annual Meeting, Montreal, Quebec, Canada}
(\byear{2019}).
\burl{https://archive.ismrm.org/2019/4819.html}
\end{bchapter}
\endbibitem

\end{thebibliography}
